\pdfoutput=1
\documentclass[PAPER, atlasdraft=false, cernpreprint, texlive=2023, UKenglish, texmf, orcidlogo]{atlasdoc}
 
\usepackage{atlaspackage}
\usepackage{atlasbiblatex}
 
\usepackage{atlasphysics}

\graphicspath{{logos/}{figures/}}
 
\usepackage{ANA-STDM-2021-05-PAPER-defs}
\usepackage{multirow}

\AtlasTitle{Evidence of pair production of longitudinally polarised vector bosons and study of CP properties in $ZZ \to 4\ell$ events with the ATLAS detector at $\sqrt{s} = 13$~\TeV}

\AtlasAbstract{
A study of the polarisation and CP properties in $ZZ$ production is presented.
The used data set corresponds to an integrated luminosity of \lumi~fb$^{-1}$ of proton–proton collisions at a centre-of-mass energy of $13$~\TeV
recorded by the ATLAS detector at the Large Hadron Collider.
The $ZZ$ candidate events are reconstructed using two same-flavour opposite-charge electron or muon pairs.
The production of two longitudinally polarised $Z$ bosons is measured with a significance of 4.3 standard deviations,
and its cross-section is measured in a fiducial phase space to be \xsecObsZZLL~fb, consistent with the next-to-leading-order Standard Model prediction.
The inclusive differential cross-section as a function of a CP-sensitive angular observable is also measured.
The results are used to constrain anomalous CP-odd neutral triple gauge couplings.
}

\AtlasRefCode{STDM-2021-05}
 
\PreprintIdNumber{CERN-EP-2023-199}

\AtlasJournalRef{\JHEP 12 (2023) 107}
\AtlasDOI{10.1007/JHEP12(2023)107}

\hypersetup{pdftitle={ATLAS document},pdfauthor={The ATLAS Collaboration}}
 
\begin{document}
 
\maketitle
 
\tableofcontents

\section{Introduction}
\label{sec:intro}

\begin{figure}[h]
\centering
\null\hfill
\subfloat[][]{\label{fig:fey_qqZZ}\includegraphics[width=0.25\textwidth]{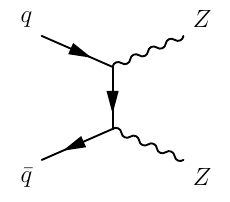}}
\hfill
\subfloat[][]{\label{fig:fey_ggZZ}\includegraphics[width=0.25\textwidth]{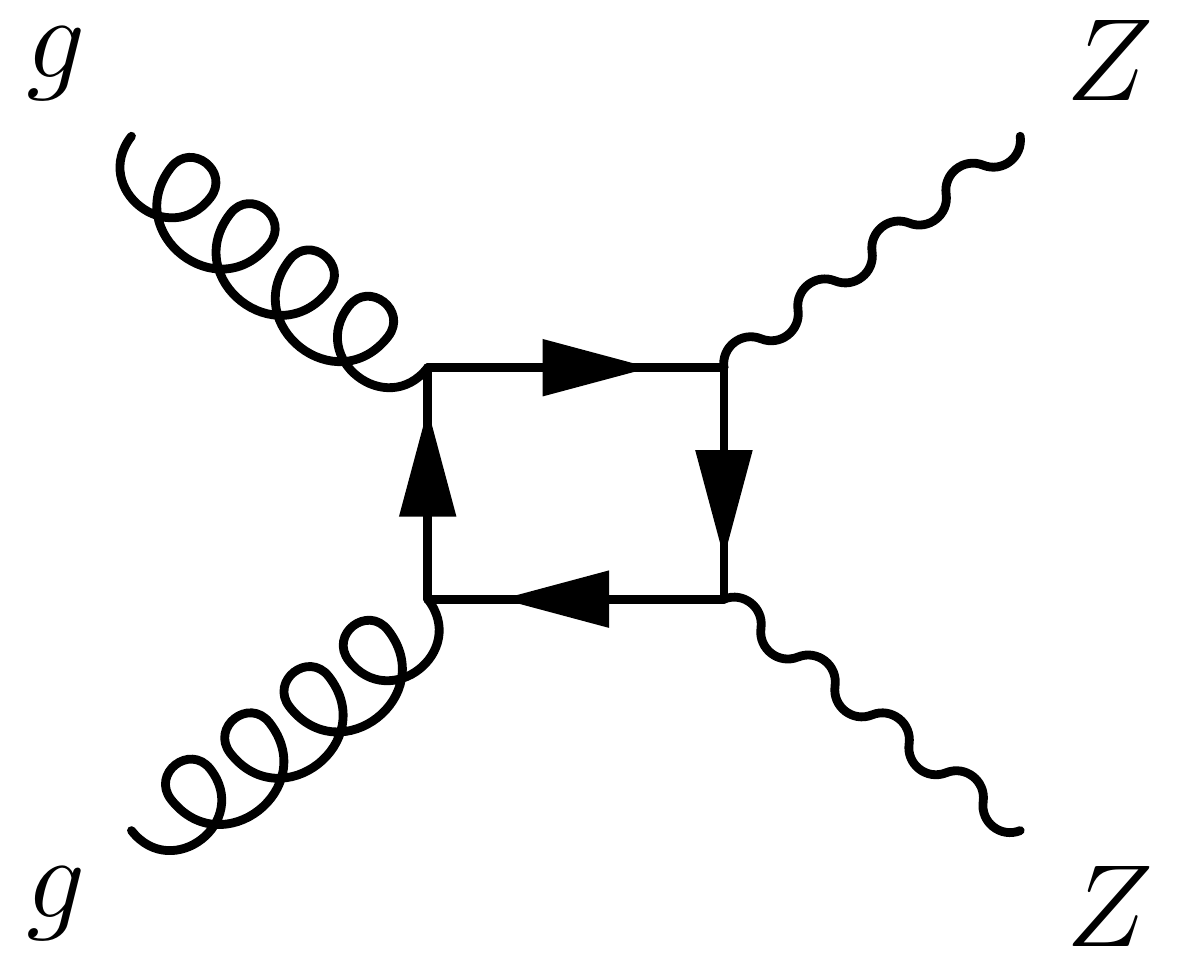}}
\hfill\null
\hfill
\subfloat[][]{\label{fig:fey_EWKqqZZ}\includegraphics[width=0.25\textwidth]{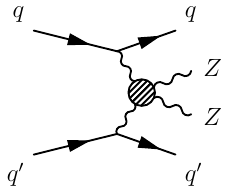}}
\hfill\null
\caption{Examples of leading-order Feynman diagrams for $ZZ$ production in $pp$ collisions: (a) $q\bar{q}$-initiated, (b) $gg$-initiated, and (c) EW $qq \rightarrow ZZjj$. The internal fermion lines are quarks.}
\label{fig:fey_ZZ}
\end{figure}

Measurements of $Z$ boson pair~($ZZ$) production in proton--proton~($pp$) collisions at the Large Hadron Collider~(LHC)
provide an important test of the Standard Model~(SM) gauge structure in the electroweak~(EW) sector.
Representative leading-order Feynman diagrams of $ZZ$ production at the LHC are shown in Figure~\ref{fig:fey_ZZ}.
With increased luminosity, experimental data enables studies beyond precision measurements of integrated and differential cross-sections,
such as the weak boson polarisation and charge conjugation~(C) and parity~(P) properties.
The polarisation measurement of massive weak bosons is a direct probe of the Electroweak Symmetry Breaking mechanism,
through which the $W$ and $Z$ bosons obtain their longitudinally polarised states.
Diboson polarisation measurements, especially those probing longitudinally polarised vector bosons,
provide unique sensitivity to physics beyond the SM~(BSM)~\cite{Liu:2018pkg}.
 
Polarisation measurements performed with LHC data have  mainly focused on single bosons such as in $W$-boson production~\cite{CMS-EWK-10-014, STDM-2011-37},
$Z$-boson production~\cite{STDM-2014-10,CMS-SMP-13-010}, and $W$ bosons from top-quark decays~\cite{TOPQ-2016-02,CMS-TOP-13-008,TOPQ-2018-02}.
Single-boson polarisation states have also been measured in $WZ$ production by the ATLAS~\cite{STDM-2018-03} and CMS~\cite{CMS-SMP-20-014} Collaborations.
Recently first measurements on joint-polarisation states of weak bosons have also been reported.
The production of two transversely polarised $W$ bosons has been measured by the CMS Collaboration in EW production of the same-sign $W^{\pm}W^{\pm}$ boson pairs~\cite{CMS-SMP-20-006}, while the the ATLAS Collaboration has measured the joint-polarisation states of longitudinally or transversely polarised bosons in inclusive $WZ$ production~\cite{STDM-2022-01}.
In the latter, the fraction of diboson events with a simultaneous longitudinal polarisation~(LL) was observed with a significance of $7.1$ standard deviations.

This paper presents a measurement of the production of two longitudinally polarised $Z$ bosons~(\ZLZL) in the decay channel $ZZ\to \fourl$,
where $\ell$ and $\ell^{\prime}$ can be an electron or a muon.
The $Z$-boson candidates are reconstructed with same-flavour, opposite-charge (SFOC) electron or muon pairs, and they are required to be on-shell with $|\mll - \mZ|<10$~\GeV,
where \mll\ is the invariant mass of the lepton pair and \mZ\ is the $Z$-boson pole mass~\cite{ParticleDataGroup:2022pth}.
 
The violation of CP symmetries is required to explain the matter-antimatter asymmetry in the universe, and it is well known that there is insufficient CP violation in the SM~\cite{Gavela:1993ts, Cohen:1993nk, Huet:1994jb}.
The measurement of CP-sensitive observables in diboson production can be utilised to explore new sources of CP violation in the gauge-boson sector.
CP-violating effects in weak-boson self-interactions were studied in various measurements of diboson production at the LHC by constraining the CP-odd anomalous neutral triple gauge couplings~(aNTGC), including those entering the $ZZZ$ and $ZZ\gamma$ vertexes, using $ZZ$ production in ATLAS and CMS \cite{STDM-2012-02,STDM-2014-16,STDM-2017-03,STDM-2016-15,CMS-SMP-13-005-fixed,CMS-SMP-12-016,CMS-SMP-19-001}.
Such experimental searches primarily derive constraints on anomalous triple gauge boson couplings~(aTGC) using event rates or cross-section measurements without employing dedicated CP-sensitive observables.
 
This paper presents the differential cross-section for a dedicated CP-odd angular observable, referred to as the \textit{Optimal Observable}~(OO). The OO is defined in Section~\ref{sec:meth} using the decay products of weak bosons in $ZZ$ production, in such a way as to be sensitive to BSM amplitudes through the interference to the SM~\cite{Rahaman:2016pqj, Rahaman:2018ujg}. The results are then reinterpreted to constrain aNTGC using an effective vertex function approach~\cite{ATGCZZ}. The ATLAS Collaboration has previously used such type of dedicated CP-sensitive observables in the EW $Zjj$ production to test CP violation in the weak-boson self-interactions~\cite{STDM-2017-27}.
 
The CP property is studied using an aNTGC vertex that can be parameterised with two coupling parameters \ffZ\ and \ffg\ that violate the CP symmetry. By using such parameters, the cross-section in any given bin of the CP-sensitive observable can be parameterised as
\begin{equation}
\sigma^{i} = \sigma^i_{\mathrm{SM}} + c\cdot \sigma^{i}_{\mathrm{interference}} + c^2 \cdot \sigma^{i}_{\mathrm{quadratic}}, \label{eq:xs_bsm}
\end{equation}
where the superscript $i$ is the bin index of the CP-sensitive observable, $c$ is the CP-odd aNTGC, $\sigma^i_{\mathrm{SM}}$ is the prediction from the SM, $\sigma^{i}_{\mathrm{interference}}$ is the linear interference between the SM and the aNTGC, and $\sigma^{i}_{\mathrm{quadratic}}$ is the quadratic contribution of the aNTGC.
As pointed out in Ref.~\cite{aNTGC}, for the aNTGC, the quadratic term dominates over the linear interference term.
Existing constraints on a CP-odd aNTGC stem primarily from their effect on the cross-section in the high-\pT\ regime, derived using kinematic observables such as the leading $\pT^{Z}$.
Such high-\pT\ sensitive kinematic observables cannot distinguish CP-even and CP-odd effects.
This paper presents a search for CP violation using the unfolded OO which is sensitive to the interference terms.


\section{ATLAS detector}
\label{sec:detector}

\newcommand{\AtlasCoordFootnote}{
ATLAS uses a right-handed coordinate system with its origin at the nominal interaction point (IP)
in the centre of the detector and the \(z\)-axis along the beam pipe.
The \(x\)-axis points from the IP to the centre of the LHC ring,
and the \(y\)-axis points upwards.
Cylindrical coordinates \((r,\phi)\) are used in the transverse plane,
\(\phi\) being the azimuthal angle around the \(z\)-axis.
The pseudorapidity is defined in terms of the polar angle \(\theta\) as \(\eta = -\ln \tan(\theta/2)\).
Angular distance is measured in units of \(\Delta R \equiv \sqrt{(\Delta\eta)^{2} + (\Delta\phi)^{2}}\).}

The ATLAS experiment~\cite{PERF-2007-01} at the LHC is a multipurpose particle detector
with a forward--backward symmetric cylindrical geometry and a near \(4\pi\) coverage in
solid angle.\footnote{\AtlasCoordFootnote}
It consists of an inner tracking detector~(ID) surrounded by a thin superconducting solenoid
providing a \qty{2}{\tesla} axial magnetic field, electromagnetic and hadron calorimeters, and a muon spectrometer~(MS).
The inner tracking detector covers the pseudorapidity range \(|\eta| < 2.5\).
It consists of silicon pixel, silicon microstrip, and transition radiation tracking detectors.
Lead/liquid-argon (LAr) sampling calorimeters provide electromagnetic (EM) energy measurements
with high granularity.
A steel/scintillator-tile hadron calorimeter covers the central pseudorapidity range (\(|\eta| < 1.7\)).
The endcap and forward regions are instrumented with LAr calorimeters
for both the EM and hadronic energy measurements up to \(|\eta| = 4.9\).
The muon spectrometer surrounds the calorimeters and is based on
three large superconducting air-core toroidal magnets with eight coils each.
The field integral of the toroids ranges between \num{2.0} and \qty{6.0}{\tesla\metre}
across most of the detector.
The muon spectrometer includes a system of precision tracking chambers and fast detectors for triggering.
A two-level trigger system is used to select events.
The first-level trigger is implemented in hardware and uses a subset of the detector information
to accept events at a rate below \qty{100}{\kHz}.
This is followed by a software-based trigger that
reduces the accepted event rate to \qty{1}{\kHz} on average
depending on the data-taking conditions.
An extensive software suite~\cite{ATL-SOFT-PUB-2021-001} is used in data simulation, in the reconstruction
and analysis of real and simulated data, in detector operations, and in the trigger and data acquisition
systems of the experiment.


\section{Data and simulation}
\label{sec:samp}

The data collected by the ATLAS experiment during the $2015$-$2018$ data-taking period of the LHC at a centre-of-mass energy of $\sqrt{s}= 13$~\TeV is analysed,
and corresponds to an integrated luminosity of \lumi~\ifb.
The data events are triggered using single-lepton and dilepton triggers with the minimum trigger threshold, depending on data-taking periods, varying between $20-26$~\GeV for single-lepton triggers and $8-24$~\GeV for dilepton triggers~\cite{ElectronTriggers_Run2, MuonTriggers_Run2}. The trigger efficiency is nearly 100\% for signal events after offline event selections.
 
Simulations of the signal SM processes of the on-shell production of two $Z$ bosons and background processes resulting in two SFOC lepton pairs are derived using Monte Carlo (MC) generators. The \qqtoZZ\ (Figure~\ref{fig:fey_qqZZ}) process was modelled using the \Sherpa $2.2.2$ generator~\cite{qqZZ_Sherpa}.
The matrix elements~(ME) were calculated at next-to-leading-order~(NLO) accuracy in QCD for up to one additional parton emission and leading-order~(LO) accuracy for up to three additional parton emissions. The matrix element calculations were matched and merged with the \Sherpa parton shower based on the Catani--Seymour dipole factorisation~\cite{Comix, CS_DipoleShower}, using the MEPS@NLO prescription~\cite{NLOPS_Matching, MEPS_Matching_Sherpa, MEPS_Matching_Sherpa_NLO}. The virtual QCD corrections were provided by the OpenLoops library~\cite{OpenLoops1, OpenLoops2, OpenLoops3}. The \NNPDF[3.0nnlo] set of parton distribution functions (PDF)~\cite{Ball:2014uwa} and the dedicated set of tuned parton-shower parameters developed by the \Sherpa authors were used for generating this sample.
The higher-order corrections from NLO electroweak effects were taken into account for the \qqtoZZ\ sample by applying reweighting corrections as a function of the four-lepton invariant mass $m_{4\ell}$~\cite{EWkCorrectionqqZZ, EWkCorrectionqqZZ_2017}. 
 
An alternative prediction of the \qqtoZZ\ process with \POWHEGBOX v2~\cite{Alioli:2010xd, Melia:2011tj, Nason:2013ydw} is used for estimating the theory modelling uncertainty. This sample was generated at NLO accuracy in QCD and interfaced with \PYTHIA[8.186]~\cite{Sjostrand:2007gs} for modelling the parton shower, hadronisation, and effect of underlying events using the AZNLO set of tuned parameters~(`tune')~\cite{AZNLO}. For the hard-scattering and parton showering, CT10 PDF~\cite{Lai:2010vv} and CTEQ6L1 PDF~\cite{Pumplin:2002vw} sets were used, respectively. A higher-order correction as a function of $m_{4\ell}$ was obtained using a \textsc{Matrix} NNLO QCD prediction~\cite{POWHEGNLO, POWHEGNLO_1, POWHEGNLO_2, POWHEGNLO_3} and applied to these events. The correction was defined as the ratio of the cross-section at NNLO QCD accuracy to the one at NLO QCD accuracy.
 
The loop-induced \ggtoZZ\ process~(Figure~\ref{fig:fey_ggZZ}) was also modelled with the \Sherpa $2.2.2$ generator using ME calculated with LO accuracy for up to one additional parton emission. As detailed in Ref.~\cite{STDM-2018-30}, the higher-order QCD effects are accounted for by normalising the LO prediction to NNLO using NLO~\cite{ggZZQCDCorr,ggZZQCDCorr_1} and NNLO~\cite{Passarino:2013bha,deFlorian:2016spz} predictions.
The same parton showering, matching, and merging schemes as for the \Sherpa\ \qqtoZZ\ simulations were used.
The EW production of $ZZ$ in association with two jets, \qqtoZZjj ~(Figure~\ref{fig:fey_EWKqqZZ}), was modelled by \MGNLO[2.6.7]~\cite{Alwall:2014hca}. The matrix element was calculated at LO in QCD, and the \NNPDF[3.0nnlo] PDF set was used.
\PYTHIA[8.244]~\cite{Sjostrand:2014zea} was used to simulate parton showering, hadronisation, and underlying-event activity using the \NNPDF[2.3lo]~\cite{Ball:2012cx} PDF set and the A14 tune~\cite{ATL-PHYS-PUB-2014-021}.
The three processes, \qqtoZZ, \ggtoZZ, and \qqtoZZjj, collectively simulate the total signal events for this measurement.
 
In addition to the simulations for the inclusive $ZZ$ production, samples for different polarisation states were also produced.
The polarised \qqtoZZ\ signal samples were simulated using \MGNLO[2.7.3]~\cite{Alwall:2014hca, BuarqueFranzosi:2019boy},
with the ME calculated at LO in perturbative QCD and with the \NNPDF[3.0nlo]~\cite{Ball:2014uwa} PDF set. The events were interfaced to \PYTHIA[8.240] to model the parton shower, hadronisation, and underlying event, with parameters set according to the A14 tune and using the \NNPDF[2.3lo] set of PDFs.
Polarised samples were simulated corresponding to the three helicity states, \ZTZT\ for two transversely polarised $Z$ bosons, \ZTZL\ for one transversely polarised $Z$ boson and one longitudinally polarised $Z$ boson, and \ZLZL\ for two longitudinally polarised $Z$ bosons, respectively, with each $Z$ boson decaying independently into an electron or a muon pair. The $Z$-boson polarisation is defined in the centre-of-mass (CM) frame of the two $Z$ bosons, which is a natural choice for diboson production~\cite{Denner:2021csi}.
To account for the real part of the NLO corrections, the QCD-induced events were simulated with up to two jets in the matrix element at LO and merged with \PYTHIA[8] parton showers using the CKKW-L scheme~\cite{Lonnblad:2001iq, Lonnblad:2011xx}.
The EW production of the polarised $ZZ$ samples was similarly generated with \MGNLO[2.7.3], and they are used together with the QCD-induced \qqtoZZ\ samples in the analysis.
Generation of polarised MC events for the loop-induced \ggtoZZ\ process is not possible,
and a MC event reweighting procedure is applied for the modelling as described in detail in Section~\ref{sec:meth_pol}.
The off-shell $Z/\gamma^{*}$ contribution is included in the inclusive $ZZ$ samples but not in the polarised simulations, and the impact is found to be negligible after the event selection with a requirement of two on-shell $Z$ bosons, as described in Section~\ref{subsec:EventSel}.
 
Background events consist of four-lepton events originating from $t\bar{t}Z$ and $VVZ$ processes, as well as four-lepton events containing at least one non-prompt lepton. The $t\bar{t}Z$ events were modelled by the \Sherpa $2.2.0$ generator at LO accuracy with up to one additional parton emission using the MEPS@LO setup~\cite{MEPS_Matching_Sherpa, MEPS_Matching_Sherpa_NLO} and the \NNPDF[3.0nnlo] PDF set. The sample was scaled to reproduce the prior $t\bar{t}Z$ cross-section measurement from ATLAS~\cite{ttV}. Events from the $VVZ$ processes were simulated by the Sherpa $2.2.2$ generator with NLO accuracy in QCD for the inclusive process and LO accuracy in QCD for up to two additional parton emissions. The \NNPDF[3.0nnlo] PDF set and the default \Sherpa parton showering scheme were used. The prediction was scaled to match the prior measurement of the triboson production from ATLAS~\cite{VVV}.
 
The events with any of the four leptons originating from non-prompt sources are estimated by using a semi-data-driven method discussed in Section~\ref{sec:bkg}. The method uses information about the origin of the non-prompt leptons from simulations of the relevant processes. The primary source of this background is $WZ$ production with the leptonic decays of vector bosons and was modelled with \Sherpa $2.2.2$. The same setup and parameters as with the \qqtoZZ\ modelling were used. The events with non-prompt leptons arising from \Zjets\ processes were modelled using \Sherpa $2.2.1$ at NLO accuracy in QCD for up to two parton emission and LO accuracy for up to four parton emission. The ME were calculated using the \textsc{Comix} and \textsc{OpenLoops} libraries~\cite{Comix}. Matching and merging were performed with the \Sherpa parton shower scheme using the MEPS@NLO prescription. The \NNPDF[3.0nnlo] PDF set was used, and the events are normalised to a prediction accurate to NNLO in QCD~\cite{WZNLO} for the inclusive production.
The non-prompt events originating from the $t\bar{t}$ processes were modelled using the \POWHEGBOX v2 generator at NLO accuracy and the \NNPDF[3.0nlo] PDF set. The prediction was interfaced with \PYTHIA[8.230] for modelling parton showering, hadronisation, and underlying event with parameters set according to the A14 tune.
 
The MC samples used in the BSM aNTGC interpretation of the CP study were generated at LO in QCD with \MGNLO[2.7.2] using a UFO model~\cite{UFO} as implemented in Ref.~\cite{aNTGC}, and interfaced with \PYTHIA[8.244] for the modelling of parton shower, hadronisation, and underlying event. The aNTGC predictions are reweighted by applying a per-bin $k$-factor, derived by comparing the \Sherpa\ prediction to the LO \MGNLO \qqtoZZ\ prediction in each bin of the OO to account for the missing higher-order effects in the BSM prediction.
 
The predictions from the MC simulations were passed through a detailed simulation of the ATLAS detector~\cite{ATLAS_SimInf} based on \GEANT~\cite{Agostinelli:2002hh}. The effect of multiple $pp$ interactions in the same bunch crossing, known as pile--up, was emulated by overlaying inelastic $pp$ collisions, simulated with \Pythia[8.186] using the \NNPDF[2.3lo] set of PDFs and the A3 tune~\cite{ATL-PHYS-PUB-2016-017}. The events were then reweighted to match the distribution of the average number of interactions per bunch crossing observed in the data for different data-taking periods. The simulated events were reconstructed using the same algorithms used for the data. Additionally, efficiencies for the trigger, lepton reconstruction, identification, and isolation in MC events were corrected to match those measured in the data.


\section{Fiducial region, object and event selections}
\label{sec:selec}

\subsection{Fiducial region definition}
\label{subsec:FidDef}
The fiducial phase space is defined close to the detector's kinematic acceptance using particle-level prompt\footnote{\textit{Prompt objects} are leptons and photons that do not originate from hadrons.} leptons produced by a MC event generator, without simulating the effects of the detector. 
The particle-level prompt leptons are \textit{dressed} by adding the four-momenta of nearby prompt photons within a small cone of $\Delta R < 0.1$. The dressed electrons (muons) are required to be within the detector's acceptance such that they satisfy $\pT > 7~(5)$~GeV and $|\eta| < 2.47 ~(2.7)$. The fiducial phase space of the analysis does not include the negligible contribution of prompt leptons from $\tau$-lepton decays. Events must have a minimum of four prompt dressed leptons to be grouped into at least two SFOC lepton pairs, and the leading and sub-leading leptons must have $\pT > 20$~GeV. The angular separation between any two leptons is required to satisfy $\Delta R > 0.05$ to reduce the double counting of detector signatures while keeping leptons from possible boosted production scenarios. The invariant mass of any SFOC lepton pair is required to satisfy $m_{\ell\ell} > 5$~GeV.
 
The two SFOC lepton pairs whose invariant masses are closest and next closest to  \mZ are designated as an event quadruplet. In the on-shell $ZZ$ region, the resolution on \mZ is comparable to the mass difference between the two $Z$ bosons, resulting in some events with inconsistent definitions of the leading and sub-leading pairs at particle and detector levels. This increases the resolution-induced bin migrations that need to be corrected for by the unfolding procedure. Therefore, once the quadruplet is formed, the leading~(sub-leading)-lepton pair $Z_{1}$ ($Z_{2}$) is identified as the one with the larger (smaller) value of absolute rapidity, i.e., $|y_{\ell\ell}|$. Based on these requirements, the events are divided into three categories, $4e$ events with two $e^{+}e^{-}$ pairs, $4\mu$ events with two $\mu^{+}\mu^{-}$ pairs and $2e2\mu$ events where one of the pairs is $e^{+}e^{-}$ and the other is $\mu^{+}\mu^{-}$. The invariant mass of each SFOC lepton pair is required to be within $|m_{\ell\ell} - \mZ | < 10$~GeV, and the invariant mass of the four leptons is required to be $m_{4\ell}>180$~GeV, motivated to select only on-shell $ZZ$ events.
 
\subsection{Object selection}
\label{subsec:ObjSel}
Events are required to contain at least one reconstructed $pp$ collision vertex candidate with at least two associated ID tracks with $\pT > 0.5$~\GeV.
The vertex with the largest sum of $\pT^{2}$ of tracks is considered to be the primary interaction vertex.
 
Electrons are reconstructed by matching the topological energy clusters deposited in the electromagnetic calorimeters to the tracks in the ID~\cite{EGAM-2018-01}. The electron identification is based on a multivariate-likelihood technique that takes information about clusters' shower shapes in the electromagnetic calorimeters, ID track properties, and the quality of track-cluster matching.
\textit{Baseline} electrons, used for the non-prompt background estimate, are required to satisfy $\pT > 7$~GeV, $|\eta| < 2.47 $ and the `VeryLoose' identification criteria~\cite{EGAM-2018-01} and loose association with the primary hard-scatter vertex by requiring $|z_{0}\sin\theta|<0.5$~mm, where $z_0$ is the longitudinal impact parameter and $\theta$ is the polar angle of the track.
\textit{Signal} electrons that define signal events are required to satisfy all of the baseline electron criteria and the stricter `LooseBLayer' identification criteria and transverse impact parameter significance of $|d_{0}|/\sigma_{d_{0}} <  5 $, where $d_0$ is the transverse impact parameter relative to the beamline, and $\sigma_{d_{0}}$ is its uncertainty.
Prompt leptons originating from hard scattering are characterised by low activity around them in the $\eta-\phi$ plane.
Therefore, a signal electron must be isolated from other particles by applying criteria for the \pT-dependent isolation variable, which is defined using the electron's track and calorimeter energy deposits~\cite{EGAM-2018-01}.
 
Muons are identified using information from various parts of the detector, the ID, the MS, and the calorimeters. A `Loose' identification working point~\cite{MuonReconstruction_Run2} is used. ID tracks identified as muons based on their calorimetric energy deposits or the presence of individual muon segments are included in the region $|\eta|< 0.1$, and the stand-alone MS tracks are added in the region $2.5 < |\eta| < 2.7$.
\textit{Baseline} muons used in the non-prompt background estimate are required to satisfy the `Loose' identification criteria, $\pT > 5$~GeV, $|\eta|<2.7$ and loose track-to-vertex association of $|z_{0}\sin\theta| < 0.5$~mm. \textit{Signal} muons are required to satisfy all criteria for baseline muons and transverse impact parameter significance of $|d_{0}|/\sigma_{d_{0}} < 3 $. Similarly to the signal electrons, signal muons are required to satisfy additional isolation criteria on the \pT-dependent isolation variable, defined using the tracks and particle-flow objects used in muon reconstruction~\cite{MuonReconstruction_Run2}.
 
Jets are reconstructed from particle-flow objects~\cite{Jets_Pflow} using the anti-$k_{T}$ algorithm with a radius parameter of $R=0.4$~\cite{ATLAS_antiKT,Fastjet}.
The jet-energy scale is calibrated using simulation and further corrected with in situ methods~\cite{JETM-2018-05}.
Candidate jets are required to satisfy $\pT > 15$~GeV and $|\eta|<4.5$, and are used in the non-prompt background estimation studies.
A jet-vertex tagger~\cite{PERF-2014-03} is applied to jets with $\pT < 60$~\GeV and $|\eta| < 2.4$ to suppress jets that originate from pile-up.
In addition, $b$-jets are identified using a multivariate $b$-tagging algorithm~\cite{FTAG-2019-07}. The chosen $b$-tagging algorithm has an efficiency of $85\%$ for $b$-jets and a rejection factor of $33$ against light-flavour jets, measured in \ttbar\ events~\cite{FTAG-2018-01}.
 
An overlap removal is applied to avoid double counting of the detector signal, favouring leptons with higher \pT\ and preferring non-calorimeter tagged muons over electrons if they share an ID track. The overlap removal rejects any jets within $\Delta R < 0.2$ of an electron or jets associated with less than three ID tracks if they overlap with a muon.
 
\subsection{Event selection}
\label{subsec:EventSel}
 
The reconstructed events must have at least four baseline leptons, and the leading and sub-leading leptons must satisfy $\pT > 20$~GeV. In each event, all possible SFOC lepton pairs are formed by requiring $m_{\ell \ell} > 5$~GeV and $\Delta R_{\ell\ell} > 0.05$ to suppress contributions from the leptonic decays of resonance hadrons. Like the fiducial region, a quadruplet is formed from two SFOC pairs with the lowest values of $|m_{\ell\ell} - \mZ|$. The SFOC lepton pair with the largest value of $|y_{\ell\ell}|$ is defined as the leading $Z$ boson candidate. To ensure the selection of on-shell $ZZ$ events, the invariant masses of both SFOC lepton pairs are required to be $|m_{\ell\ell}- \mZ| < 10 $~GeV and the invariant mass of the quadruplet is required to satisfy $m_{4\ell} > 180$~GeV. Each lepton of the quadruplet is required to satisfy the signal lepton definition for the signal region selection. Events with either one or more lepton in a quadruplet failing to meet the signal lepton requirement are used in non-prompt background estimate. Due to the on-shell requirement on both $Z$ bosons, in $4e$ and $4\mu$ final states there are less than $2\%$ of events with mispaired leptons, which has no significant impact for both the CP and polarisation measurements.


\FloatBarrier
\section{Background estimation}
\label{sec:bkg}

After the event selection, the background consists of events with one or more of the reconstructed leptons in the quadruplet not originating from a $Z$ boson decay. The background processes with prompt leptons from $t\bar{t}Z$ and fully leptonic decays of triboson processes are estimated using MC simulations.
The measurement also accounts for an additional source of background from non-prompt leptons originating from hadron decays or misidentification of jets. A data-driven \textit{fake-factor} method is used to estimate the non-prompt background.
 
A \textit{fake-factor} quantity, defined as the ratio of signal leptons to the number of baseline leptons failing to meet the signal lepton criteria, is measured from data in dedicated control regions~(CR) enriched with non-prompt leptons. The CR where the fake factors are evaluated consists of events with two prompt leptons from a physics process such as \Zjets\ or $t\bar{t}$ and additional leptons from other sources such as jets. The same triggers and object-level kinematic selections as in the signal region are also applied to select the CR events. The \Zjets\ CR is selected by requiring an SFOC signal lepton pair from a $Z$-boson decay with an invariant mass of $76 < m_{\ell\ell} < 106$~\GeV and additional baseline leptons. Similar to the signal region, SFOC lepton pairs are formed with leptons having $\Delta R > 0.05$ and $m_{\ell\ell}>5$~\GeV. Additionally, events are required to have a missing transverse energy less than 50~\GeV to suppress the contamination from the $WZ$ process. The \ttbar\ CR is defined by requiring an SFOC signal lepton pair, at least one $b$-tagged jet and at least one additional baseline lepton.
 
The additional leptons in the \Zjets\ and \ttbar\ CRs originate predominantly from non-prompt sources such as heavy- or light-flavour jets and are used to derive the fake factors. The additional non-prompt leptons in \Zjets\ events arise dominantly from light-flavour jets, whereas they arise dominantly from heavy-flavour jets in \ttbar\ events. To match the heavy-flavour composition in the signal region, the two types of events are first weighted and combined. The combination weight is evaluated by comparing the fraction of the non-prompt leptons from heavy-flavour decays in the \Zjets\ and \ttbar\ CRs to that of the signal region using the simulated samples discussed in Section~\ref{sec:samp}. Since the compositions for the non-prompt electrons and muons are different in the signal region, the weights are evaluated separately to combine the control region events that have additional baseline electrons and muons.
When evaluating the fake factors, an estimate of the genuine baseline prompt leptons that fail the signal requirements from the $WZ$ MC simulation is subtracted for both the \Zjets\ and \ttbar\ events. The fake factor is measured as a function of \pT\ and $\eta$ of the non-prompt leptons and the number of jets in an event.
 
A fake-lepton enriched region is selected with a minimum of four baseline leptons passing the same kinematic selection as in the signal region, but with at least one baseline-not-signal lepton, which is defined as leptons passing the baseline selection but failing the signal-lepton requirement.
The non-prompt event yield in the signal region is then estimated by applying a fake-factor weight to each baseline-not-signal lepton.
Four-lepton events containing prompt baseline-not-signal leptons are removed from the estimate using simulations.
 
The background estimate is validated by comparing the prediction of the fake-factor method to the data in two dedicated validation regions where the non-prompt background events are expected to be dominant. The first one is the different-flavour validation region which has the same event selection requirements as the signal region but requires the leptons forming one of the pairs to have different flavours. The second one is the same-charge validation region, which requires the leptons in one of the pairs to have the same charge. In both of the validation regions, the data agree closely with the sum of the estimated non-prompt background yield and the MC prediction within the statistical uncertainties of the data.
 
Three sources of uncertainties in the background yield are considered. The first uncertainty is related to the statistical uncertainty of the non-prompt leptons in the control region used to calculate the fake factors and ranges from $10-80\%$. The second set of uncertainty, ranging from $2-40\%$, is related to the theory uncertainties of the subtracted prompt-lepton component in the control region, which are dominated by the QCD scale variations. The third and dominant uncertainty, ranging from $20-100\%$, is the statistical precision on the number of events with at least four baseline leptons and at least one failing to satisfy the signal-lepton requirement.


\section{Measurement methods}
\label{sec:meth}

\subsection{Polarisation measurements}
\label{sec:meth_pol}

The $Z$ boson can be either transversely polarised or longitudinally polarised, and the polarisation fractions depend on the transverse momentum
of the $Z$ boson~\cite{Stirling:2012zt}. These effects lead to different kinematic properties of the production and the final decay states of the $Z$-boson pair.
To extract the fraction of \ZLZL\ events from the reconstructed $ZZ$ candidate events, a multivariate technique based on a boosted decision tree~(BDT)~\cite{TMVA}
is used to enhance the separation between \ZLZL\ and \ZTZX\ (\ZTZT\ or \ZTZL) events.
After a dedicated optimisation study to maximise the \ZLZL\ signal sensitivity, the input variables used in the BDT are the following:
\costho(\costht), where \tho(\tht) is the angle between the negatively charged final-state lepton in the \Zo(\Zt) rest frame and the direction of
flight of the \Zo(\Zt) boson in the four-lepton rest frame;
\costhst, where \thst\ is the production angle of the \Zo\ defined in the four-lepton rest frame; and
\dphio(\dphit), the azimuthal separation of the two leptons from \Zo(\Zt) defined in the four-lepton rest frame.
The angles are illustrated in Figure~\ref{fig:angle_frame}.
Other kinematic variables, such as the \pT\ and rapidity of \Zo\ and \Zt\, also have substantial separation power, but they are not included in the BDT training
to reduce theoretical modelling uncertainties.

\begin{figure}[htb]
\centering
\includegraphics[width = 0.6\textwidth]{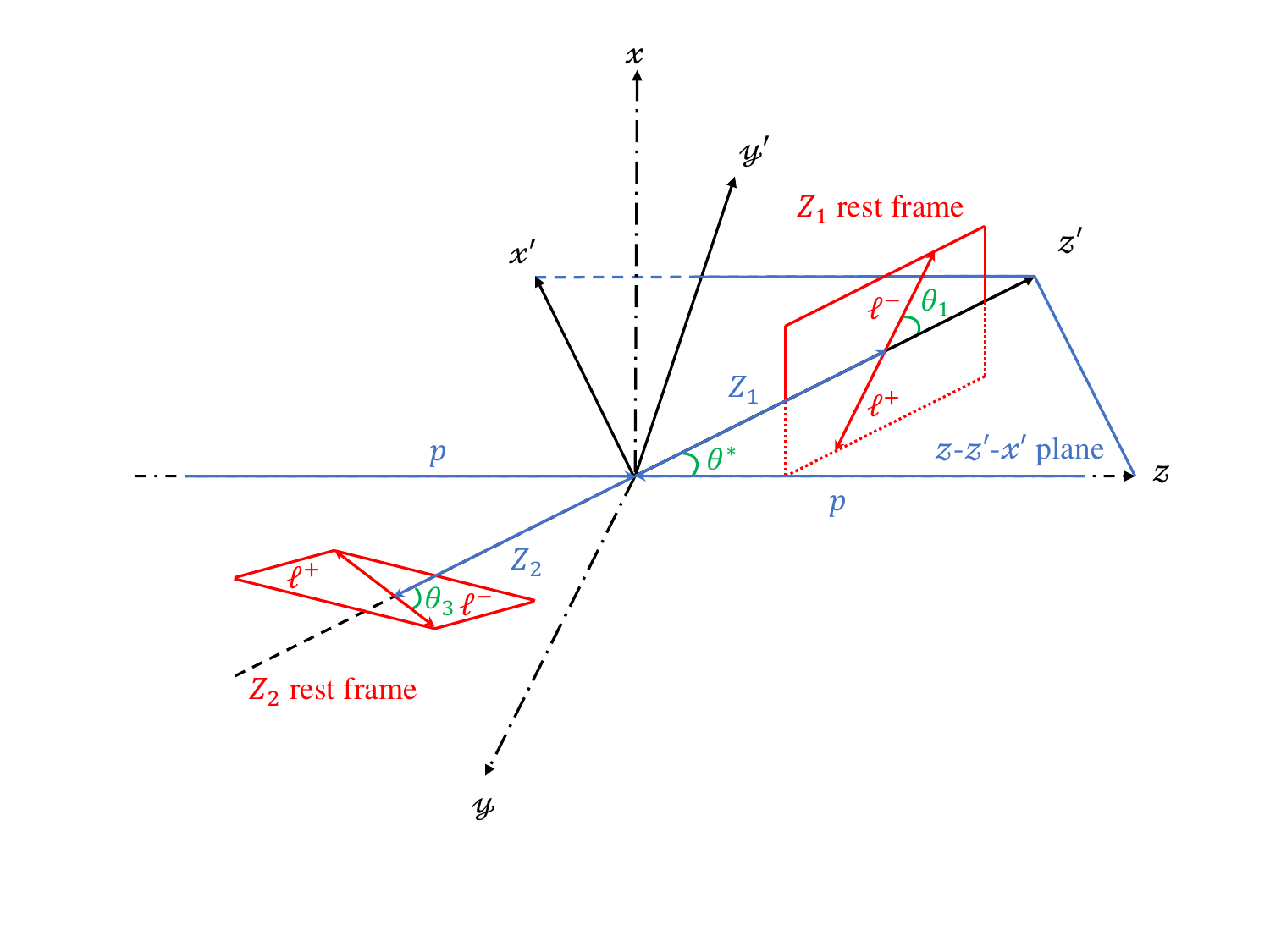}
\caption{Definition of the angles used for the polarisation measurement and the reference frame used to define the CP-sensitive angles.
The $xyz$-frame~(dot-dashed) is the laboratory frame with the $z$-axis along the beam direction.
The $x^{\prime}y^{\prime}z^{\prime}$-frame~(solid) is a new frame used to define the CP-sensitive angles.
The $z^{\prime}$-axis is defined as the direction of motion of the \Zo\ boson in the four-lepton rest frame.
The $x^{\prime}$-axis defines the reaction plane containing the laboratory $z$-axis and the $z^{\prime}$-axis. The right-hand rule gives the $y^{\prime}$-axis.}
\label{fig:angle_frame}
\end{figure}

The MC templates for different $ZZ$ polarisation states were generated at LO in QCD.
Higher-order corrections, in both the QCD and EW, on MC templates of the different $ZZ$ polarisation states have to be taken
into account when extracting the \ZLZL\ fraction from data.
Recently the combined NLO QCD and EW corrections  and the loop-induced \ggtoZZ\ corrections to the polarisation structure of $ZZ$ production
were calculated at fixed-order with the \MoCaNLO\ program~\cite{Denner:2021csi}.
Differential cross-sections for different polarisation states of the \qqtoZZ\ and \ggtoZZ\ processes were provided for several kinematic observables.
As in the simulated MC samples, the polarisation definition in the \MoCaNLO\ program is also based on the CM frame of the two $Z$ bosons.
In order to incorporate the  fixed-order higher-order corrections into the polarisation measurement, a three-step reweighting method is established,
using either a one-dimensional~(1D) observable or two-dimensional~(2D) observables.
\begin{itemize}
\item \textbf{Step 1: 1D reweighting for each individual polarisation state.} In this step, the reweighting is done separately for the \qqtoZZ\ and \ggtoZZ\ processes.
For \qqtoZZ, the combined NLO QCD and EW corrections, in a multiplicative approach, as a function of the \costho\ variable are applied by taking the ratio of the differential cross-sections
calculated from \MoCaNLO\ at NLO and the ones from the \MGNLO\ MC samples at the particle level in the fiducial phase space of the measurement,
for \ZTZT, \ZTZL\ and \ZLZL\ events, respectively. An additional reweighting as a function of \costho\ is applied to account for the missing higher-order QCD and parton shower effects by taking the ratio of inclusive \SHERPA
\qqtoZZ\ predictions at the particle level to the inclusive \MoCaNLO\ calculations.
The impact of these two 1D reweighting corrections on the BDT discriminant is mostly of about 20\% on the normalisation, and the shape variation is only of $2-4$\%.
For the \ggtoZZ\ process, which contributes to about 15\% of the total signal yield, only the inclusive \SHERPA\ MC sample is available, and the \MoCaNLO\ program provides polarised differential cross-sections at LO.
Thus the inclusive \SHERPA\ MC sample is reweighted to obtain polarised templates of \ZTZT, \ZTZL\ and \ZLZL,
by taking the fraction of polarised and inclusive cross-sections calculated by \MoCaNLO\ as a function of \costho.
No reweighting is applied to the EW \qqtoZZjj\ process and the original MC simulation of polarised samples is used.
\item \textbf{Step 2: 1D reweighting for the interference effect.} The simulated polarised samples do not consider the interference effects among different polarisation states,
while such interference effects are found to be non-negligible in some kinematic regions where the contribution could reach up to 5\%~\cite{Denner:2021csi}.
A dedicated template for the interference term is therefore constructed by reweighting the inclusive \SHERPA\ \qqtoZZ\ events with \MoCaNLO\ calculations that include interference contributions, by taking the difference between the inclusive cross-sections and the sum of the three polarised cross-sections as a function of \costho. For \ggtoZZ\ events, the interference effect is found to be negligible and thus ignored. For the subleading EW \qqtoZZjj\ process, the interference effect is not included either.
\item \textbf{Step 3: 2D reweighting for the residual higher-order corrections.} Four templates, including three polarisation
states and the interference term, are obtained after the two reweighting steps described above. A closure test is performed by comparing the sum of the four templates and the prediction given by the inclusive \SHERPA
MC events. Residual discrepancies are observed, which could be due to the non-closure of the 1D reweighting method, resulting from missing higher-order QCD and parton shower effects.
An additional 2D reweighting is applied to each of the three polarisation templates to correct the mismodelling by taking the ratio of the non-closure effect and the sum of the
three polarisation templates as a function of \costhst and \dphio.
The impact of this 2D reweighting on the BDT discriminant is mostly on the shape with a maximum variation of about 10\%.
\end{itemize}
 
Figure~\ref{fig:pol_BDT_reweighting} shows the BDT distribution of the three polarisation templates for the \qqtoZZ\ process before and after the reweighting procedure.
To extract the fraction of \ZLZL\ events, a profile binned maximum-likelihood fit~\cite{Moneta:2010pm,RooFit,Cranmer:1456844} to the
BDT distribution is performed using the final templates for the three polarisation states and the interference term, and other non-$ZZ$ background contributions.
The normalisation factors of the \ZLZL\ template, \muZZLL, referred as the signal strength, and the combined $\ZTZT+\ZTZL$ templates, \muZZTX, are allowed to float in the fit. The signal strength is the ratio of the measured signal contribution relative to the SM expectation.
A validation study is performed to show the robustness of the templates, by applying a similar fit to the inclusive \POWHEGBOX MC events.
The extracted \ZLZL\ fraction is compared with the predicted \ZLZL\ fraction, and they are found to be consistent within uncertainties.
 
\begin{figure}[h]
\centering
\includegraphics[width=0.6\textwidth]{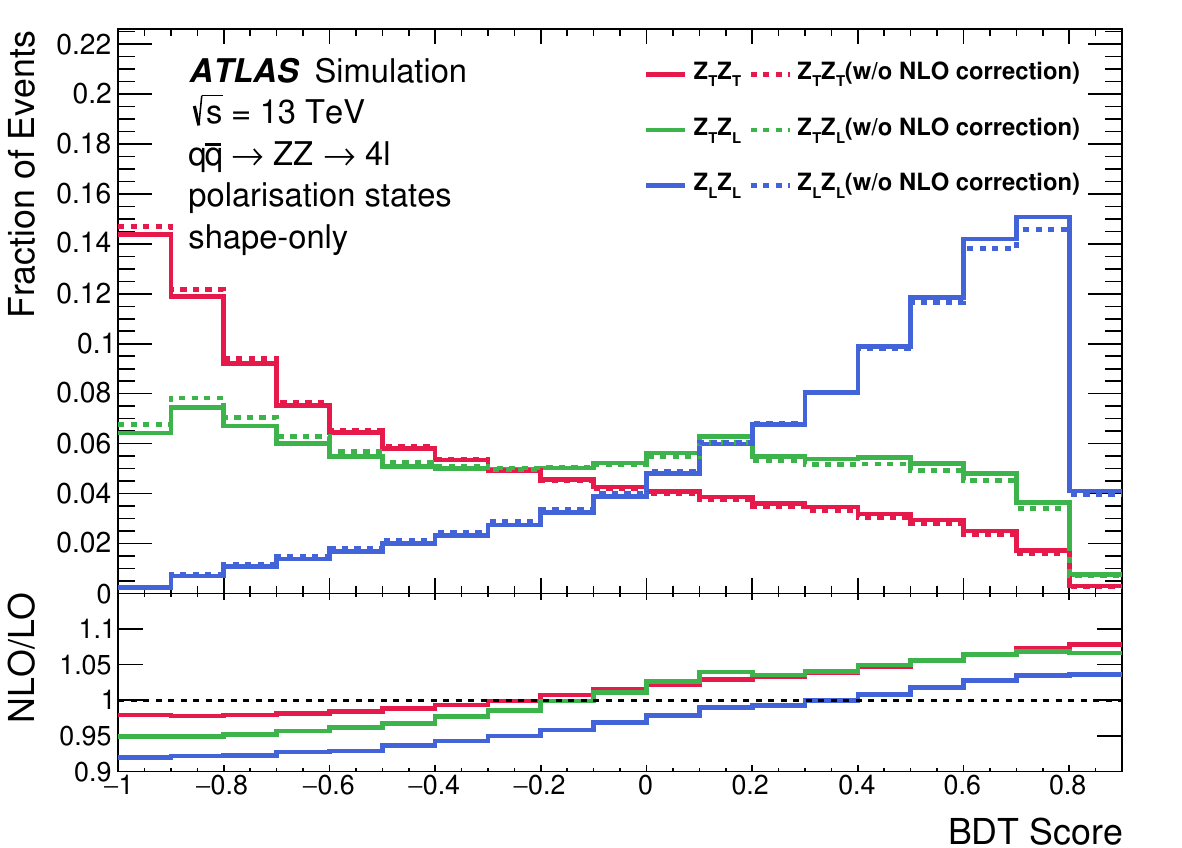}
\caption{BDT distributions of the three polarisation templates for the \qqtoZZ\ process, before~(dashed lines) and after~(solid lines) the reweighting procedure to account for higher-order corrections.
All distributions are normalised to the same area.
The lower panel shows the ratio of the templates after the corrections to those before the corrections.
}
\label{fig:pol_BDT_reweighting}
\end{figure}


\subsection{Study of CP property}
\label{sec:meth_CP}
 
\subsubsection{CP-odd Optimal Observable}
\label{sec:meth_opt_obs}

The OO defined for the CP study combines the CP-sensitive polar and azimuthal angles of both $Z$-boson systems, providing additional CP sensitivity from shape differences between the SM and aNTGC predictions. The CP-sensitive polar angles \tho(\tht) for the \Zo(\Zt) boson are already defined in Section~\ref{sec:meth_pol} and illustrated in Figure~\ref{fig:angle_frame}. The CP-sensitive azimuthal angles \phio\ and \phit\ are reconstructed in a reference frame illustrated in Figure~\ref{fig:angle_frame} that allows a direct measure of the $Z$-boson spin as discussed in Ref. \cite{Tyz_cit1, Rahaman:2016pqj}. The CP-sensitive azimuthal angle \phio(\phit) is the azimuthal angle of the negative lepton in the \Zo(\Zt) rest frame in this new axis system. The differential cross-sections for \tho(\tht) and \phio(\phit) are symmetric in the SM but asymmetric in the presence of a CP-odd aNTGC.
 
To improve the sensitivity, the two CP-sensitive angles \tho(\tht) and \phio(\phit) are combined to form an angular observable $T_{yz,1(3)} = \sin\phi_{1(3)} \times \cos\theta_{1(3)}$ that enhances the asymmetry for each $Z$-boson system. Figures~\ref{fig:SM_Tyz1_Tyz3_2d} and \ref{fig:BSM_Tyz1_Tyz3_2d} show the 2D differential distributions of the CP-sensitive observable \Tyz\ of the two $Z$ bosons, the symmetric SM prediction and the asymmetric BSM prediction in the presence of a non-zero \ffZ\ parameter, respectively. 
 
\begin{figure}[htb]
\subfloat[]{\includegraphics[width = 0.45\textwidth]{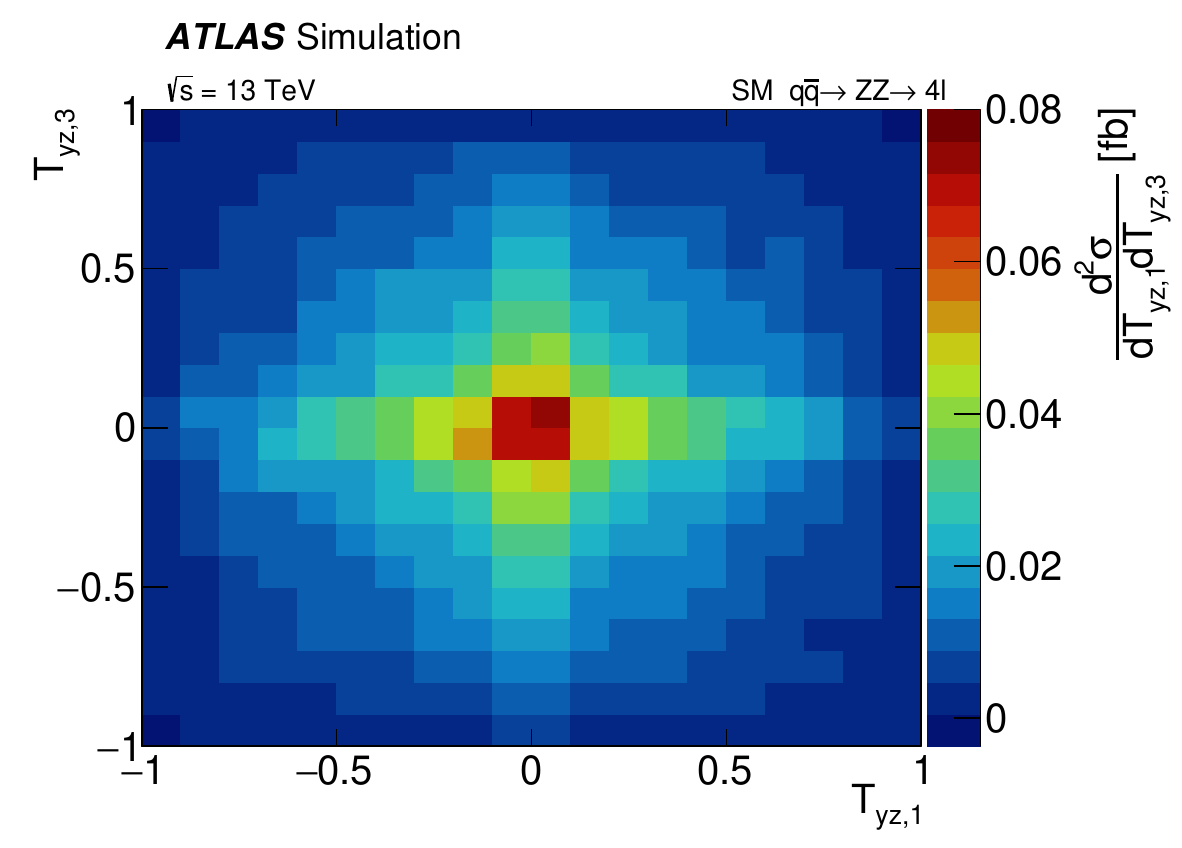}\label{fig:SM_Tyz1_Tyz3_2d}}
\subfloat[]{\includegraphics[width = 0.45\textwidth]{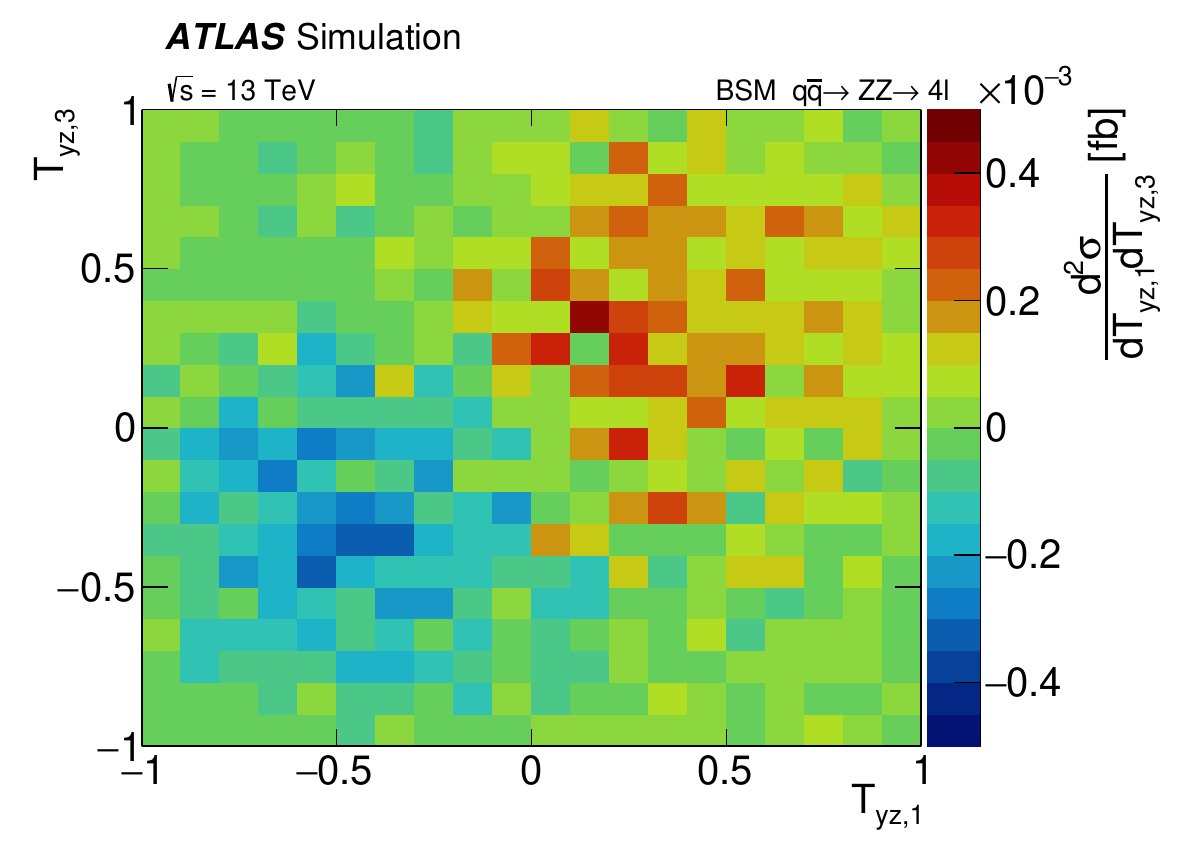}\label{fig:BSM_Tyz1_Tyz3_2d}}
\caption{ Particle level 2D differential cross-sections of \Tyz\ of the two $Z$ bosons for the $\qqtoZZ \rightarrow 4\ell$ process as predicted by (a) the SM and (b) in the presence of the BSM aNTGC vertex. The BSM prediction shows the contribution of the interference effects only, excluding the quadratic term in Equation~(\ref{eq:xs_bsm}), when $\ffZ =1$.}\label{fig:SM_Obs}
\end{figure}
 
As observed in Figure~\ref{fig:BSM_Tyz1_Tyz3_2d}, the first (top right) and the third (bottom left) quadrants where both of the $Z$ bosons have negative and positive \Tyz\ values, respectively, are the most sensitive regions of the 2D \Tyz\ distribution. The OO \OTyzTyz\ is defined from the $2D$ distribution of \Tyz\ by grouping together the sensitive and non-sensitive bins to enhance the sensitivity for the four-lepton system. Each bin of the \OTyzTyz\ observable represents approximately an L-shaped grouping of the bins around the $T_{yz,3} = T_{yz,1}$ line as shown in Figure~\ref{fig:1dmap}. 
The small fraction of events with mis-paired leptons in the $ZZ \rightarrow 4e ~(4\mu)$ final states was studied and found to have negligible impact on the CP-sensitivity of the OO.
 
Figure~\ref{fig:OO_DataMC} shows the data compared with the total SM signal and background MC prediction at the detector level of the OO \OTyzTyz. The bins $1$ to $7$ and $24$ to $30$ in Figure~\ref{fig:OO_DataMC} represent the third quadrant and the first quadrant, respectively, of the 2D distribution of \Tyzo\ vs \Tyzt\ shown in Figure~\ref{fig:SM_Obs}. In these two quadrants, the \Tyz\ observables for both $Z$ bosons have the same sign in the SM and are the most CP-sensitive region, along with the two central bins representing the bin number $15$ and $16$ of the OO.
The data agree closely with the prediction within the measurement's statistical precision and systematic uncertainties. Figure~\ref{fig:OO_DataMC} also shows an asymmetric prediction in the presence of a CP odd BSM coupling when $\ffZ =1$.
 
\begin{figure}[htb]
\centering
\subfloat[]{\includegraphics[width = 0.48\textwidth]{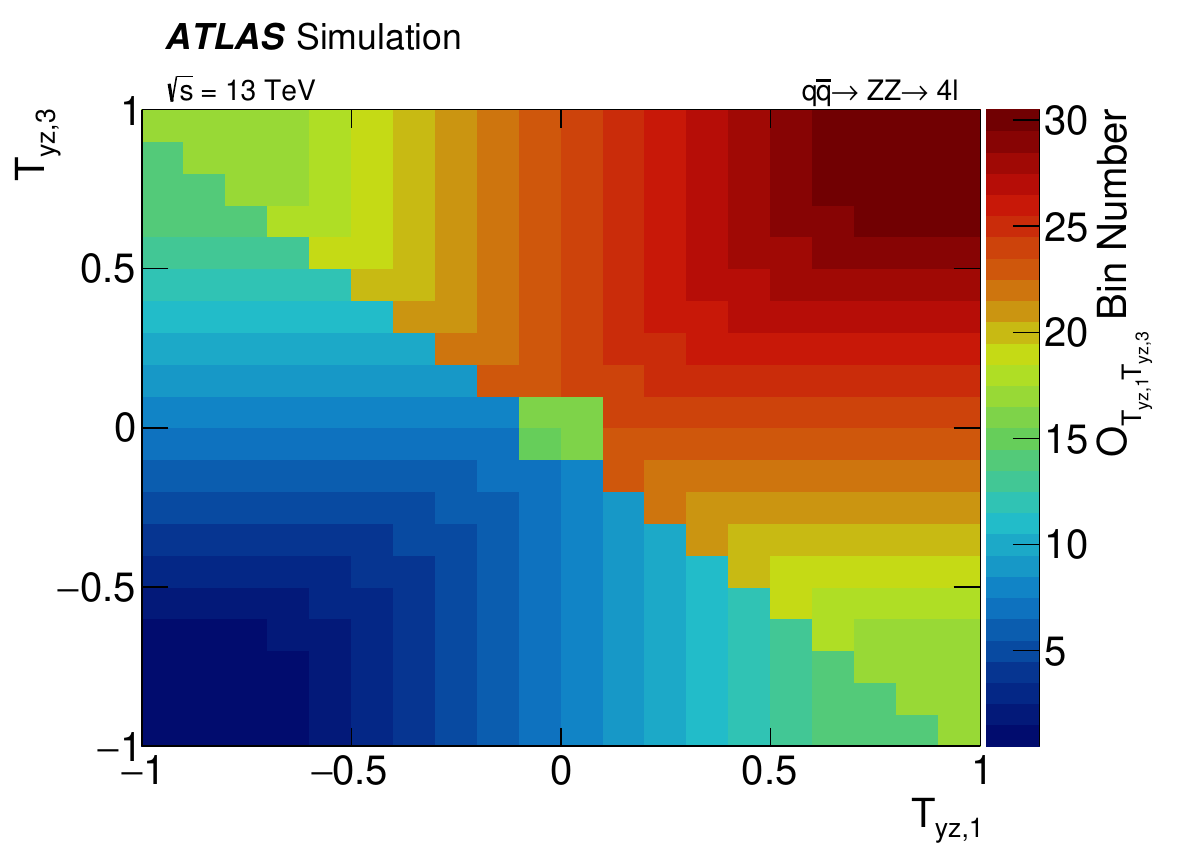}\label{fig:1dmap}}
\subfloat[]{\includegraphics[width = 0.45\textwidth]{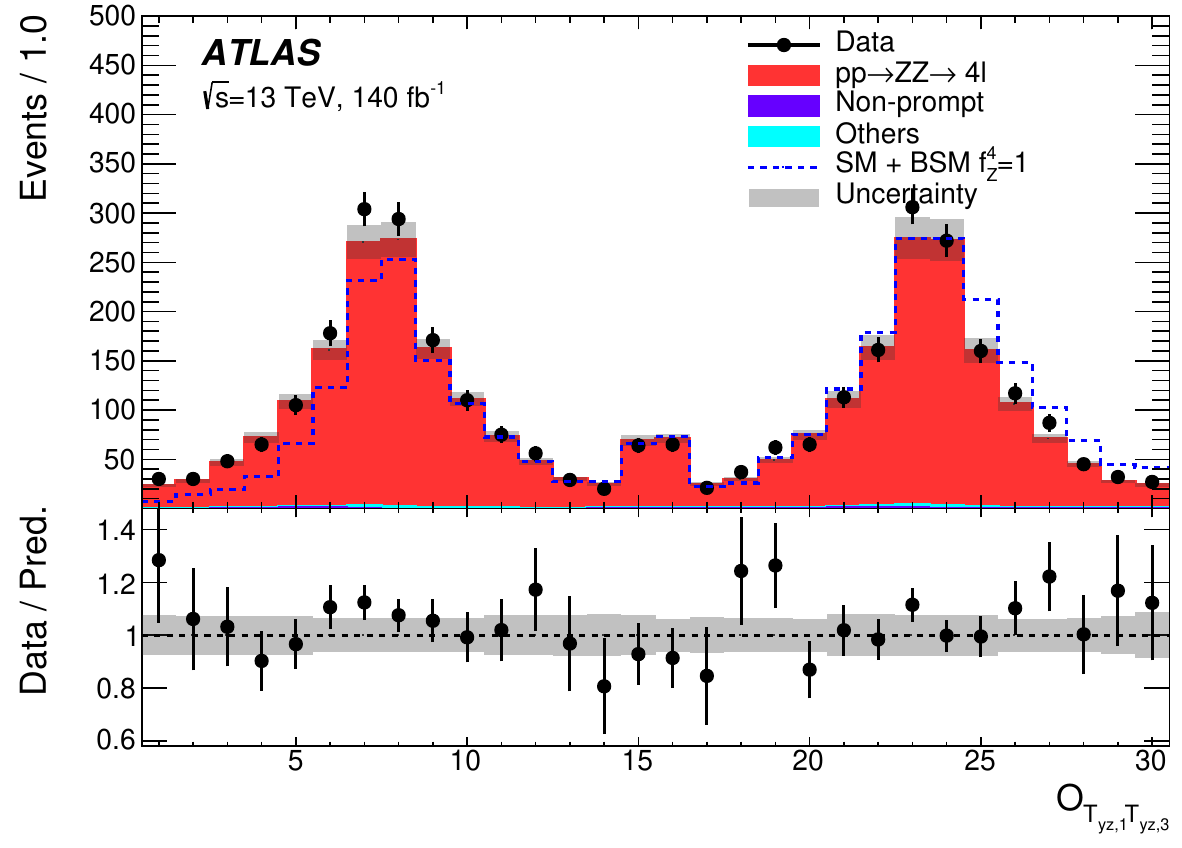}\label{fig:OO_DataMC}}
\caption{ (a) The $2D \rightarrow 1D$ mapping and (b) the detector-level measurement of the OO \OTyzTyz. The measured distribution is compared with the SM signal prediction and the total background. The `Others' category includes the contribution from $\ttbar Z$ and $VVZ$ processes. The non-prompt background is estimated by using the fake-factor method. The grey band represents the effect of the total theoretical and experimental uncertainties for the detector-level predictions, and the vertical error bars on data represent the statistical uncertainties. The effect of CP-odd BSM coupling is also represented by the dashed histogram.}\label{fig:OO_DataMC_Map}
\end{figure}


\subsubsection{Detector corrections}
\label{sec:meth_unfolding}

Particle-level differential cross-sections for the on-shell $ZZ$ production are obtained by correcting the detector effects such as inefficiency and resolution. The background-subtracted event yields are corrected using an iterative Bayesian unfolding method~\cite{IBU}.
 
The first step of the correction multiplies each bin yield by a fiducial correction factor obtained from the \Sherpa\ SM prediction, which accounts for the events that satisfy the detector level but fail to satisfy the fiducial-level event selections. This correction accounts for the $5-20\%$ of the fake fiducial events in various bins caused by the resolution effects.
Then, the detector resolution-induced bin migrations are corrected iteratively using the SM particle-level distribution as the initial prior. With an increasing number of iterations, the statistical uncertainty increases, and the residual bias relative to the prior decreases due to the improvement of its knowledge. Two iterations were deemed optimal as a compromise between the increasing statistical uncertainty and decreasing bias. The final step in the unfolding procedure is to correct for the detector inefficiency by dividing the per-bin yield by the ratio of the number of events satisfying both the particle- and reconstruction-level selections to the number of events passing the particle-level selections.
 
A data-driven closure test is performed to evaluate the model dependence of the unfolding method. This test first simulates a pseudo-data sample by reweighting the SM prediction to the shape observed in the data. The pseudo-data sample is then unfolded using the nominal SM prediction. The comparison of the unfolded pseudo-data with the reweighted particle-level prediction gives the intrinsic bias of the unfolding method, which was found to be less than $1\%$ in each bin of the unfolded \OTyzTyz\ observable in the case of two iterations of unfolding. This resulting bias is taken as a systematic uncertainty of the final result. Moreover, the uncertainty related to the choice of the generator in the unfolding is studied using the alternative \POWHEG\ prediction of the \qqtoZZ\ process, which is reweighted to match the nominal \Sherpa\ lineshape to avoid double counting of the data-driven bias. The generator bias estimated by comparing the difference between the unfolded results is negligible.
 
Additionally, an injection test is performed to evaluate the robustness of the unfolding algorithm in the presence of BSM physics in the data. A detector-level distribution for the BSM aNTGC parameter $f_{4}^{Z}=1$ is injected into the SM detector-level prediction. The BSM-injected detector level distribution is then unfolded using the inputs from the nominal SM prediction. When compared, the unfolded distribution agrees closely with the corresponding particle-level distribution within uncertainties.



\section{Systematic uncertainties}
\label{sec:syst}

Both the polarisation and CP property studies presented here are affected by some common sources of theoretical, experimental, and  background-related uncertainties.
 
\textbf{Theoretical systematic uncertainties} affecting the results of this analysis arise from two sources: uncertainties related to the QCD scale dependence and uncertainties related to the choice of the PDF set. The impact of each is propagated to the final results. The QCD scale dependency is evaluated individually for three physics processes \qqtoZZ, \ggtoZZ, and the EW \qqtoZZjj\ by varying the default choice of renormalisation and factorisation scales independently by factors of two and one-half, and then removing combinations where the variations differ by a factor of four. For each process, the envelope of the effects from these variations is taken as the scale uncertainty.
 
The PDF-related uncertainty for each of the signal samples is estimated using the PDF4LHC recommendation~\cite{Butterworth:2015oua}. The PDF variations include a set of $100$ replica variations on the nominal NNPDF set, two additional variations from the alternative PDF sets of MMHT2014nnlo~\cite{MMHT} and CT14nnlo~\cite{CT14}, and variations of the strong coupling constant by $\pm0.001$ around the nominal value of $\alpha_{S}=0.118$. The total PDF uncertainty is the absolute envelope of standard deviations of $100$ internal variations and of the two alternate PDF variations, added in quadrature with the envelope of the $\alpha_{S}$ variations.
 
For the polarisation measurements, dedicated systematic uncertainties are considered in the modelling of the polarisation templates,
including theoretical uncertainties from higher-order corrections, PDFs and $\alpha_{S}$ described above,
and the uncertainties associated with the reweighting methods as described in Section~\ref{sec:meth_pol}.
For both the \qqtoZZ\ and \ggtoZZ\ polarisation templates, the theoretical uncertainties from QCD scales are taken from the \MoCaNLO calculations
by varying the QCD scales as described above,
while for the template of the interference term, which is reweighted from the inclusive \SHERPA\ \qqtoZZ\ samples, these theoretical uncertainties are
estimated from the \SHERPA\ sample.
The parton showering and hadronisation uncertainty is estimated for the signal by comparing the nominal \PYTHIA[8] parton showering with the
alternative \textsc{Herwig}~7~\cite{Bellm:2015jjp, Bahr:2008pv} algorithm.
Uncertainties from the NLO EW corrections are estimated by taking the difference between the additive and the multiplicative prescription in \MoCaNLO\ \cite{Denner:2021csi}.
For the 1D reweighting method in Section~\ref{sec:meth_pol}, the uncertainty is estimated by comparing the reweighted templates using the nominal observable and
an alternative observable: the rapidity difference of the two $Z$ bosons.
For the additional 2D reweighting, the residual difference between the sum of the four templates and the inclusive prediction from the \SHERPA\ sample
is taken as an uncertainty.
 
For the CP study, additional uncertainties associated with the higher-order corrections applied to the inclusive \qqtoZZ\ and \ggtoZZ\ samples are considered. For the virtual NLO electroweak effects on the \qqtoZZ\ process, a 100\% uncertainty is assigned to the reweighting function to account for non-factorisable effects in events with high QCD activity~\cite{EWkCorrectionqqZZ_unc}. The uncertainty in the NLO QCD $k$-factors for the \ggtoZZ\ process is evaluated differentially as a function of the $m_{4\ell}$ as discussed in Ref.~\cite{ggZZQCDCorr_1}.
 
The uncertainties associated with reconstruction, identification, isolation and track-to-vertex matching efficiencies, and momentum resolution and scale of the leptons are the dominant \textbf{experimental uncertainties} and originate from imperfect modelling in the simulation and uncertainties in the determination of the correction factors. The uncertainties associated with each scale factor that is applied in the simulation are estimated by modifying the nominal values by their associated uncertainties~\cite{PERF-2017-01,MuonReconstruction_Run2}.
 
An uncertainty of $\pm 0.83\%$ from the measurement of Run~2 data sample luminosity~\cite{Lumi} is propagated to the final results. A systematic uncertainty related to the pile-up reweighting is included to cover the uncertainty in the ratio of the predicted and measured $pp$ inelastic cross-sections~\cite{STDM-2015-05}.

The \textbf{background-related uncertainties} are from two distinct sources, uncertainties related to the non-prompt background estimate, discussed in Section~\ref{sec:bkg}, and the uncertainty related to the background containing four prompt leptons simulated from MC\@. The simulation of $t\bar{t}Z$ and triboson background processes are normalised to ATLAS measurements, as outlined in Section~\ref{sec:samp}, and the uncertainty related is estimated by varying the normalisation of the simulated samples by the experimental precision of the ATLAS measurements.
 
For the CP study, the systematic uncertainties are propagated to the particle level through the unfolding method. The effect of each systematic source is evaluated using simulation. Since the theory variation applied affects both the detector and the particle-level yields, the resulting theory uncertainties on the unfolded cross-sections are minor. Another source of systematic uncertainty related to the intrinsic unfolding bias discussed in Section~\ref{sec:meth_unfolding} is also included in the differential cross-sections and the BSM interpretation.


\FloatBarrier
\section{Results}
\label{sec:result}

\subsection{Polarisation measurements}
\label{sec:res_pol}

The profile likelihood fit procedure described in Section~\ref{sec:meth_pol} is performed on the BDT distribution of the data.
Systematic uncertainties are modelled as Gaussian-constrained nuisance parameters in the likelihood.
Figure~\ref{fig:pol_bdt_postfit} shows the post-fit BDT distribution in the signal region.
The corresponding pre-fit and post-fit yields in the signal region are detailed in Table~\ref{tab:pol_yields}.
The decrease of the yield uncertainty for \ZTZT\ is mainly due to the constraint on the normalisation from the fit and the increase
of the yield uncertainty for \ZLZL\ is mainly due to the statistical uncertainty of the signal strength after the fit.
The \ZLZL\ signal strength is measured to be $\muZZLL = \valBkObsmuZZLL = \valObsmuZZLL$,
where the uncertainties are either statistical~(stat.) or of systematic~(syst.) nature.
This corresponds to a significance of $\sigObsmuZZLL\sigma$.
Figure~\ref{fig:pol_NLL} shows the profile likelihood ratio~\cite{Cowan:2010js} as a function of \muZZLL.
The measured results are consistent with the SM expectation of $\muZZLL = \valExpmuZZLL$ with an expected significance of $\sigExpmuZZLL\sigma$.
The normalisation factor for the \ZTZX\ template is measured to be $\muZZTX = \valObsmuZZTX$, also consistent with the SM prediction.

\begin{figure}[h]
\centering
\includegraphics[width=0.6\textwidth]{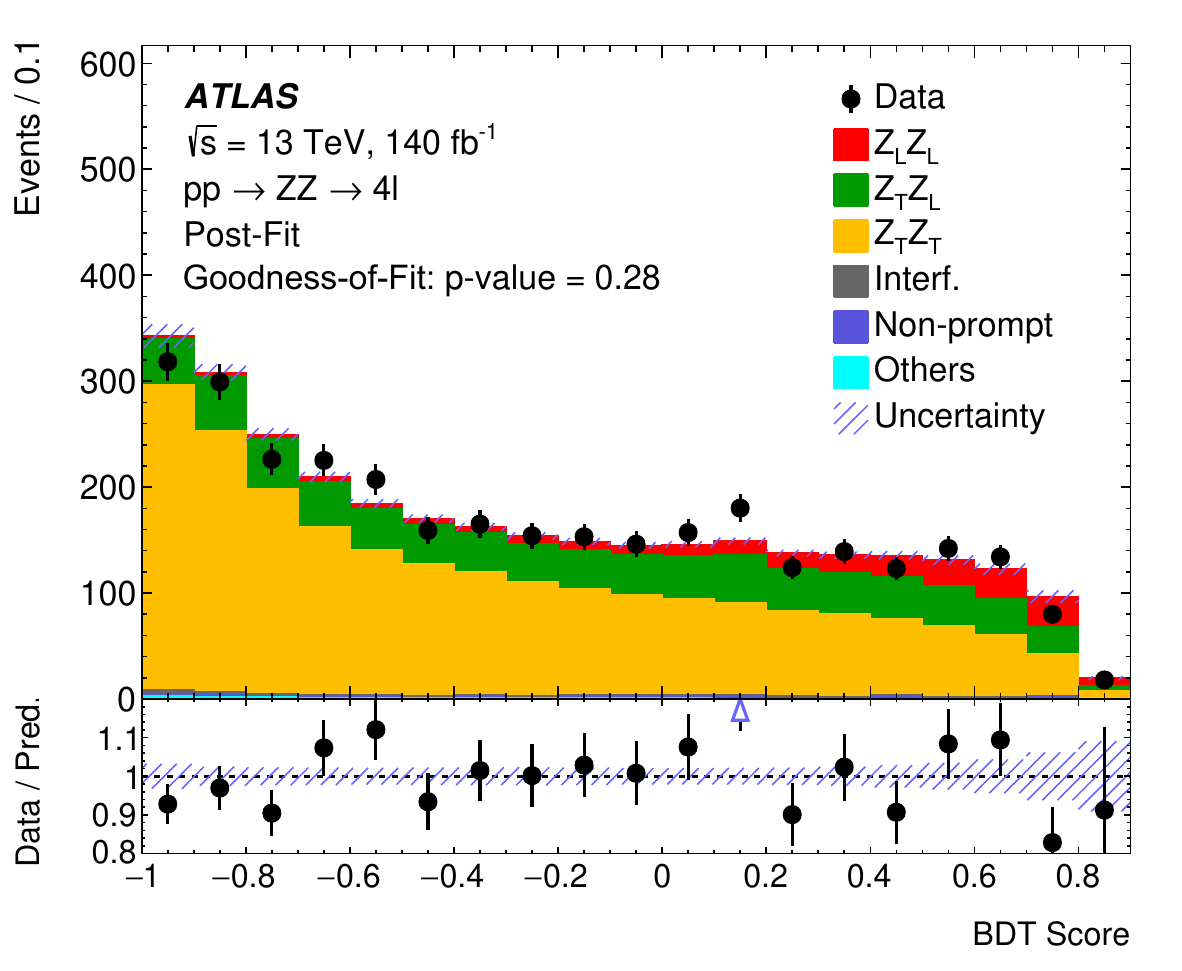}
\caption{The BDT distribution of the data and the post-fit SM predictions.
The `Others' category represents the contribution from $\ttbar Z$ and $VVZ$.
The lower panel shows the ratio of the data points to the post-fit total prediction.
The arrow indicates that the ratio lies outside the range covered by the vertical axis.
The uncertainty bands include both of the statistical and systematic uncertainties as obtained by the fit.
}
\label{fig:pol_bdt_postfit}
\end{figure}
 
\begin{figure}[h]
\centering
\includegraphics[width=0.6\textwidth]{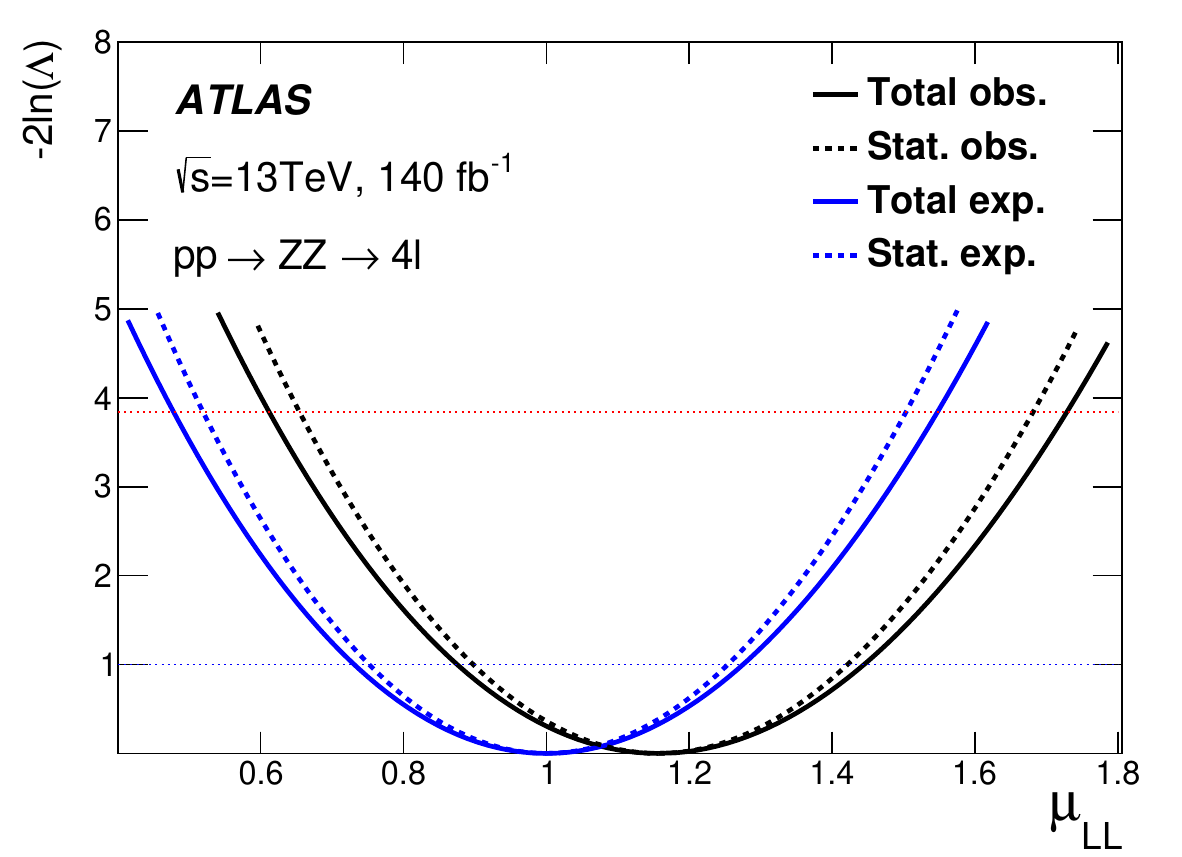}
\caption{Observed and expected profile likelihood ratio, $-2\mathrm{ln}\Lambda$, as a function of the signal strength \muZZLL,
with~(solid curves) and without~(dashed curves) systematic uncertainties included.}
\label{fig:pol_NLL}
\end{figure}

\begin{table}[!htbp]
\caption{
Expected and observed numbers of events in the signal region.
Numbers are presented before and after the fit to the BDT distribution.
The `Others' category represents the contribution from $\ttbar Z$ and $VVZ$.
For the \ZLZL\ signal, the pre-fit yield values correspond to the theoretical prediction and corresponding uncertainties.
The uncertainties include both of the statistical and systematic contributions. The uncertainty in the total yield can be
smaller than the quadrature sum of the contributions because of correlations resulting from the fit.
}
\label{tab:pol_yields}
\centering
\begin{tabular}{l l r@{\,}l!{$\pm$}r@{\,}l r@{\,}l!{$\pm$}r@{\,}l}
\toprule
\toprule
& & \multicolumn{4}{c}{Pre-fit} & \multicolumn{4}{c}{Post-fit} \\
\midrule
\multirow{4}{*}{$ZZ$} & \ZLZL\                & 189.3& & 8.7& & 220& & 54& \\
& \ZTZL\                & 710& & 29& & 711& & 29& \\
& \ZTZT\                & 2170& & 120& & 2147& & 60& \\
& Interference          & 33.7& & 2.8& & 33.4& & 2.7& \\
\midrule
\multicolumn{2}{c}{Non-prompt}          & 18.7& & 7.1& & 18.5& & 7.0& \\
\midrule
\multicolumn{2}{c}{Others}              & 20.0& & 3.7& & 19.9& & 3.7& \\
\midrule
\multicolumn{2}{c}{Total}            & 3140& & 150& & 3149& & 57& \\
\midrule
\multicolumn{2}{c}{Data}             & \multicolumn{4}{c}{3149} & \multicolumn{4}{c}{3149} \\
\bottomrule
\bottomrule
\end{tabular}
\end{table}

An additional profile likelihood fit is performed to convert the measured \muZZLL\ to the measured fiducial cross-section,
with the normalisation effects from the theoretical uncertainties of the \ZLZL\ template removed.
The fiducial cross-section of the \ZLZL\ production is measured to be $\sigma^{\mathrm{obs.}}_{\ZLZL} = \xsecBkObsZZLL\,\mathrm{fb} = \xsecObsZZLL\,\mathrm{fb}$,
consistent with the SM prediction of $\sigma^{\mathrm{pred.}}_{\ZLZL} = \xsecExpZZLL$~fb.
The SM prediction includes NLO QCD and EW corrections for the \qqtoZZ\ process, the LO prediction for the \ggtoZZ\ process, both of which are calculated from \MoCaNLO,
and the LO prediction for the EW \qqtoZZjj\ process. The uncertainty on the prediction is dominated by QCD scale and PDF uncertainties.
The measurement is limited by data statistical uncertainty and the impact of uncertainties in the measured \ZLZL\ fiducial cross-section is shown in Table~\ref{tab:table_ranking_groups},
with the leading contribution from the theoretical modelling of the polarisation templates.

\begin{table*}[htb!]
\centering
\caption{
Impact of uncertainties in the measured fiducial cross-section $\sigma_{\ZLZL}$\ from the fit.
The impact from a group of nuisance parameters is defined via quadrature subtraction of the $\sigma_{\ZLZL}$\ uncertainties: $\sqrt{(\Delta \sigma)^2 - (\Delta{\sigma^{\prime}})^2}$,
where $\Delta \sigma$ is the uncertainty of the $\sigma_{\ZLZL}$\ from the nominal fit and the $\Delta \sigma^{\prime}$ is the uncertainty of the $\sigma_{\ZLZL}$\
when this group of nuisance parameters are fixed to their best-fit values from the nominal fit.
}
\label{tab:table_ranking_groups}
\begin{tabular}{lr}
\toprule
\toprule
Contribution                                       & Relative uncertainty $[\%]$ \tabularnewline
Total                                              & \ensuremath{24}    \tabularnewline
\midrule
Data statistical uncertainty                                  & \ensuremath{23}     \tabularnewline
Total systematic uncertainty                                  & \ensuremath{8.8}     \tabularnewline
\midrule
MC statistical uncertainty                                      & \ensuremath{1.7}     \tabularnewline
Theoretical systematic uncertainties                          & \ensuremath{}      \tabularnewline
\hspace*{4mm} \qqtoZZ\ interference modelling                 & \ensuremath{6.9}      \tabularnewline
\hspace*{4mm} NLO reweighting observable choice for \qqtoZZ   & \ensuremath{3.7}      \tabularnewline
\hspace*{4mm} PDF, $\alpha_s$ and parton shower for \qqtoZZ   & \ensuremath{2.2}      \tabularnewline
\hspace*{4mm} NLO reweighting non-closure                     & \ensuremath{1.0}      \tabularnewline
\hspace*{4mm} QCD scale for \qqtoZZ                           & \ensuremath{0.2}      \tabularnewline
\hspace*{4mm} NLO EW corrections for \qqtoZZ                  & \ensuremath{0.2}      \tabularnewline
\hspace*{4mm} \ggtoZZ\ modelling                              & \ensuremath{1.4}      \tabularnewline
Experimental systematic uncertainties              & \ensuremath{}      \tabularnewline
\hspace*{4mm} Luminosity 			         & \ensuremath{0.8}      \tabularnewline
\hspace*{4mm} Muons			 		 & \ensuremath{0.6}      \tabularnewline
\hspace*{4mm} Electrons			         & \ensuremath{0.4}      \tabularnewline
\hspace*{4mm} Non-prompt background                    & \ensuremath{0.3}      \tabularnewline
\hspace*{4mm} Pile-up reweighting			 & \ensuremath{0.3}      \tabularnewline
\hspace*{4mm} Triboson and $\ttbar Z$ normalisations   & \ensuremath{0.1}      \tabularnewline
\bottomrule
\bottomrule
\end{tabular}
\end{table*}


\subsection{Unfolded differential measurement}
\label{sec:res_unfolding}

The differential cross-section for the $ZZ\rightarrow 4\ell$ production as a function of the OO \OTyzTyz\ is shown in Figure~\ref{fig:Unfolded1DMap} and compared with two different SM predictions, where the \qqtoZZ\ contribution is predicted either using the \Sherpa\ or \POWHEG\ generators.
The band represents the total uncertainties, which are small and flat, related to the theoretical, experimental, and unfolding procedure-related uncertainties on the unfolded differential cross-section. Similarly, for the SM predictions, the uncertainty bands represent the total theoretical uncertainties on the total prediction. For each bin, the measured cross-section from the data agrees closely with both sets of predicted cross-sections.
 
\begin{figure}[htb]
\centering
\includegraphics[width = 0.55\textwidth]{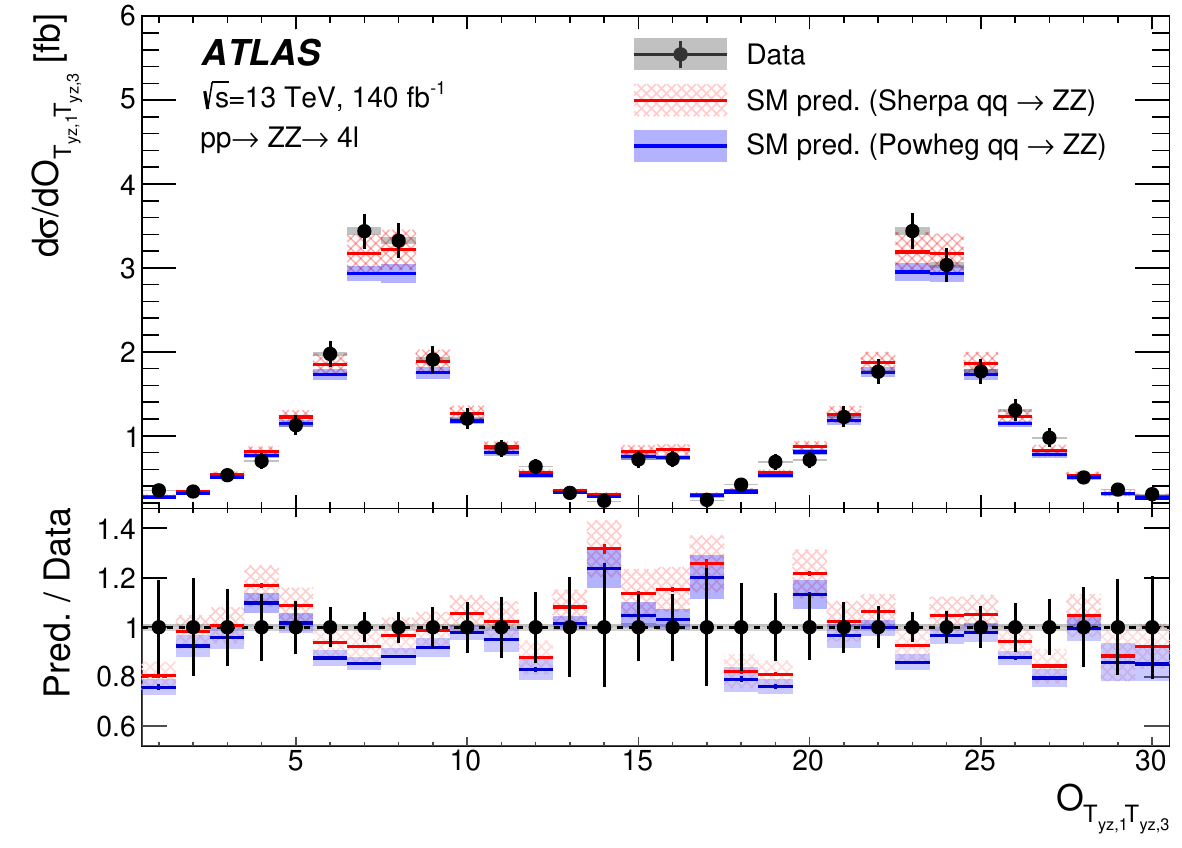}
\caption{Unfolded differential cross-section as a function of the Optimal Observable \OTyzTyz. The grey, red and blue uncertainty bands represent the respective systematic uncertainties in the measured unfolded cross-section and predicted particle-level cross-sections, where the \qqtoZZ\ contribution is predicted using either the \Sherpa\ or \POWHEG\ generators. The vertical error bars represent the total uncertainties in the measured differential cross-section.}
\label{fig:Unfolded1DMap}
\end{figure}


\subsection{BSM interpretation}
\label{sec:res_eft}

The differential cross-section as a function of \OTyzTyz\ and the corresponding $\mathrm{SM}+\mathrm{aNTGC}$ prediction are used to define a likelihood function,
\begin{equation}\label{eq:likelihoodEFT}
\mathcal{L} = \frac{1}{\sqrt{ \left(2\pi\right)^k | V |}} \exp \left\{ - \frac 1 2 \left[ \sigdata - \sigpred\left(\vec{\theta}\right) \right]^T V^{-1} \left[  \sigdata - \sigpred\left(\vec{\theta}\right)  \right] \right\}  \times \prod_{i} \mathcal{G}\left(\theta_{i}, 0, 1\right)  \, ,
\end{equation}
where \sigdata{} and $\sigpred$ are the measured and predicted differential cross-sections as a function of \OTyzTyz, respectively,
and $V$ is the total covariance matrix defined by the sum of the statistical and systematic
covariances in \sigdata{} and \sigpred.
Each systematic uncertainty is treated as fully correlated across bins but uncorrelated with other uncertainty sources. Each source of theoretical uncertainty affecting the SM particle level prediction of the \qqtoZZ, \ggtoZZ\ and EW \qqtoZZjj\ processes is implemented as nuisance parameters, $\vec{\theta}$, with Gaussian constraints, $\mathcal{G}\left(\theta_{i}, 0, 1\right)$.
 
Limits on the two CP-odd aNTGC parameters can be constrained in two scenarios, either including or excluding the quadratic term of Equation~\ref{eq:xs_bsm}. The main focus of the measurement is the linear interference-only terms, which results in an asymmetric differential cross-section. However, the difference in limits between the linear only and $\mathrm{linear}+\mathrm{quadratic}$ terms indicates the role played by the neglected operators at higher dimension in the effective Lagrangian~\cite{aNTGC}.
Thus, limits are set in these two scenarios for each CP-odd aNTGC parameter under test when the other is set to zero. The test statistics are constructed based on the profile likelihood ratio~\cite{Cowan:2010js},
\begin{equation}
q=-2\ln\frac{\mathcal{L}(c,\hat{\hat{\vec{\theta}}}(c))}{\mathcal{L}(\hat{c},\hat{\vec{\theta}}(c))} \, ,
\label{eqn:PL}
\end{equation}
where $c$ is the aNTGC parameter, $\hat{c}$ and $\hat{\vec\theta}$ are the unconditional maximum-likelihood estimators, and $\hat{\hat{\vec{\theta}}}(c)$ is the conditional maximum-likelihood estimator under the $c$ hypothesis.
 
Using the test statistic in Equation~(\ref{eqn:PL}),
which is assumed to be distributed according to a $\chi^2$ distribution with one degree of freedom following Wilks' theorem~\cite{WilksTheorem},
a 95\% confidence level for each aNTGC parameter is calculated, as shown in Table~\ref{tab:EFTLimits}.
Theoretical uncertainties related to the QCD scale, PDF, $\alpha_{S}$, and parton showering for the \qqtoZZ\ predictions have the largest impact on the estimated confidence interval.
 
From Table~\ref{tab:EFTLimits}, the constraints on the CP-odd \ffZ\ and \ffg\ aNTGC parameters using the linear interference term when the quadratic term is set to zero are looser than those obtained
using high-\pT\ kinematic observables with quadratic terms included~\cite{CMS-SMP-19-001}, as expected.
However, these are the first constraints on CP-odd aNTGC parameters using only linear interference terms with a dedicated CP-sensitive OO, based on angular observables.
Additional likelihood fits are performed with the quadratic terms included, and the constraints on \ffZ\ and \ffg\ are stronger by one order of magnitude, but since
\OTyzTyz\ is not sensitive to the high-\pT\ regime, such constraints are still not tighter than those obtained with high-\pT\ kinematic observables.
 
\begin{table*}[htb!]
\centering
\caption{ Expected and observed $95\%$ confidence interval for the two aNTGC operators using the measured particle-level differential cross-section as a function of \OTyzTyz. For the `Full' case, both the interference and the quadratic terms are included in the aNTGC prediction.}\label{tab:EFTLimits}
\begin{tabular}{c c c c c }
\toprule
\toprule
\multirow{2}{*}{aNTGC parameter} & \multicolumn{2}{c}{Interference only} & \multicolumn{2}{c}{Full} \\
& Expected & Observed & Expected & Observed \\
\midrule
\ffZ\     & [$-0.16,0.16$]      & [$-0.12,0.20$] & [$-0.013,0.012$]& [$-0.012,0.012$] \\
\midrule
\ffg\     & [$-0.30,0.30$]        & [$-0.34, 0.28$] &[$-0.015,0.015$] & [$-0.015,0.015 $] \\
\bottomrule
\bottomrule
\end{tabular}
\end{table*}



\FloatBarrier
\section{Conclusion}
\label{sec:conclusion}

Studies of the polarisation and CP properties in $ZZ \to 4\ell$ production are presented for the first time,
using proton--proton collisions at the LHC collected with the ATLAS detector at $\sqrt{s} = 13$~\TeV with an integrated luminosity of \lumi~fb$^{-1}$.
The $Z$-boson candidates are reconstructed with same-flavour, opposite-charge electron or muon pairs, and they are required to be on-shell with $|\mll - \mZ|<10$~\GeV.
The production of two simultaneously longitudinally polarised $Z$ bosons is measured with observed and expected significances of
\sigObsmuZZLL\ and \sigExpmuZZLL\ standard deviations, respectively.
The production cross-section of \ZLZL\ events is measured in a fiducial phase space as \xsecObsZZLL~fb, consistent with the Standard Model prediction
of \xsecExpZZLL~fb with NLO QCD and EW corrections considered.
The differential cross-section of the inclusive $ZZ$ boson production as a function of a CP-sensitive angular observable, \OTyzTyz, is also measured.
Furthermore, the measured differential cross-section is used to constrain the CP-odd neutral triple gauge couplings \ffZ\ and \ffg.
No significant deviations from the SM are observed.


\section*{Acknowledgements}


We thank CERN for the very successful operation of the LHC, as well as the
support staff from our institutions without whom ATLAS could not be
operated efficiently.
 
We acknowledge the support of
ANPCyT, Argentina;
YerPhI, Armenia;
ARC, Australia;
BMWFW and FWF, Austria;
ANAS, Azerbaijan;
CNPq and FAPESP, Brazil;
NSERC, NRC and CFI, Canada;
CERN;
ANID, Chile;
CAS, MOST and NSFC, China;
Minciencias, Colombia;
MEYS CR, Czech Republic;
DNRF and DNSRC, Denmark;
IN2P3-CNRS and CEA-DRF/IRFU, France;
SRNSFG, Georgia;
BMBF, HGF and MPG, Germany;
GSRI, Greece;
RGC and Hong Kong SAR, China;
ISF and Benoziyo Center, Israel;
INFN, Italy;
MEXT and JSPS, Japan;
CNRST, Morocco;
NWO, Netherlands;
RCN, Norway;
MEiN, Poland;
FCT, Portugal;
MNE/IFA, Romania;
MESTD, Serbia;
MSSR, Slovakia;
ARRS and MIZ\v{S}, Slovenia;
DSI/NRF, South Africa;
MICINN, Spain;
SRC and Wallenberg Foundation, Sweden;
SERI, SNSF and Cantons of Bern and Geneva, Switzerland;
MOST, Taiwan;
TENMAK, T\"urkiye;
STFC, United Kingdom;
DOE and NSF, United States of America.
In addition, individual groups and members have received support from
BCKDF, CANARIE, Compute Canada and CRC, Canada;
PRIMUS 21/SCI/017 and UNCE SCI/013, Czech Republic;
COST, ERC, ERDF, Horizon 2020, ICSC-NextGenerationEU and Marie Sk{\l}odowska-Curie Actions, European Union;
Investissements d'Avenir Labex, Investissements d'Avenir Idex and ANR, France;
DFG and AvH Foundation, Germany;
Herakleitos, Thales and Aristeia programmes co-financed by EU-ESF and the Greek NSRF, Greece;
BSF-NSF and MINERVA, Israel;
Norwegian Financial Mechanism 2014-2021, Norway;
NCN and NAWA, Poland;
La Caixa Banking Foundation, CERCA Programme Generalitat de Catalunya and PROMETEO and GenT Programmes Generalitat Valenciana, Spain;
G\"{o}ran Gustafssons Stiftelse, Sweden;
The Royal Society and Leverhulme Trust, United Kingdom.
 
The crucial computing support from all WLCG partners is acknowledged gratefully, in particular from CERN, the ATLAS Tier-1 facilities at TRIUMF (Canada), NDGF (Denmark, Norway, Sweden), CC-IN2P3 (France), KIT/GridKA (Germany), INFN-CNAF (Italy), NL-T1 (Netherlands), PIC (Spain), ASGC (Taiwan), RAL (UK) and BNL (USA), the Tier-2 facilities worldwide and large non-WLCG resource providers. Major contributors of computing resources are listed in Ref.~\cite{ATL-SOFT-PUB-2023-001}.


\clearpage

\printbibliography
 
\clearpage
 
\begin{flushleft}
\hypersetup{urlcolor=black}
{\Large The ATLAS Collaboration}

\bigskip

\AtlasOrcid[0000-0002-6665-4934]{G.~Aad}$^\textrm{\scriptsize 102}$,
\AtlasOrcid[0000-0002-5888-2734]{B.~Abbott}$^\textrm{\scriptsize 120}$,
\AtlasOrcid[0000-0002-1002-1652]{K.~Abeling}$^\textrm{\scriptsize 55}$,
\AtlasOrcid[0000-0001-5763-2760]{N.J.~Abicht}$^\textrm{\scriptsize 49}$,
\AtlasOrcid[0000-0002-8496-9294]{S.H.~Abidi}$^\textrm{\scriptsize 29}$,
\AtlasOrcid[0000-0002-9987-2292]{A.~Aboulhorma}$^\textrm{\scriptsize 35e}$,
\AtlasOrcid[0000-0001-5329-6640]{H.~Abramowicz}$^\textrm{\scriptsize 151}$,
\AtlasOrcid[0000-0002-1599-2896]{H.~Abreu}$^\textrm{\scriptsize 150}$,
\AtlasOrcid[0000-0003-0403-3697]{Y.~Abulaiti}$^\textrm{\scriptsize 117}$,
\AtlasOrcid[0000-0002-8588-9157]{B.S.~Acharya}$^\textrm{\scriptsize 69a,69b,m}$
% ...
\bigskip
\\

% ...
$^{29}$Physics Department, Brookhaven National Laboratory, Upton NY; United States of America.\\
% ...
$^{35}$$^{(a)}$Facult\'e des Sciences Ain Chock, R\'eseau Universitaire de Physique des Hautes Energies - Universit\'e Hassan II, Casablanca;$^{(b)}$Facult\'{e} des Sciences, Universit\'{e} Ibn-Tofail, K\'{e}nitra;$^{(c)}$Facult\'e des Sciences Semlalia, Universit\'e Cadi Ayyad, LPHEA-Marrakech;$^{(d)}$LPMR, Facult\'e des Sciences, Universit\'e Mohamed Premier, Oujda;$^{(e)}$Facult\'e des sciences, Universit\'e Mohammed V, Rabat;$^{(f)}$Institute of Applied Physics, Mohammed VI Polytechnic University, Ben Guerir; Morocco.\\
% ...
$^{49}$Fakult\"{a}t Physik , Technische Universit{\"a}t Dortmund, Dortmund; Germany.\\
% ...
$^{55}$II. Physikalisches Institut, Georg-August-Universit\"{a}t G\"ottingen, G\"ottingen; Germany.\\
% ...
$^{69}$$^{(a)}$INFN Gruppo Collegato di Udine, Sezione di Trieste, Udine;$^{(b)}$ICTP, Trieste;$^{(c)}$Dipartimento Politecnico di Ingegneria e Architettura, Universit\`a di Udine, Udine; Italy.\\
% ...
$^{102}$CPPM, Aix-Marseille Universit\'e, CNRS/IN2P3, Marseille; France.\\
% ...
$^{117}$Department of Physics, New York University, New York NY; United States of America.\\
% ...
$^{120}$Homer L. Dodge Department of Physics and Astronomy, University of Oklahoma, Norman OK; United States of America.\\
% ...
$^{150}$Department of Physics, Technion, Israel Institute of Technology, Haifa; Israel.\\
$^{151}$Raymond and Beverly Sackler School of Physics and Astronomy, Tel Aviv University, Tel Aviv; Israel.\\
% ...
$^{a}$ Also Affiliated with an institute covered by a cooperation agreement with CERN.\\
$^{b}$ Also at An-Najah National University, Nablus; Palestine.\\
$^{c}$ Also at Borough of Manhattan Community College, City University of New York, New York NY; United States of America.\\
$^{d}$ Also at Center for High Energy Physics, Peking University; China.\\
$^{e}$ Also at Center for Interdisciplinary Research and Innovation (CIRI-AUTH), Thessaloniki; Greece.\\
$^{f}$ Also at Centro Studi e Ricerche Enrico Fermi; Italy.\\
% ...
$^{m}$ Also at Department of Physics, King's College London, London; United Kingdom.\\
% ...

\end{flushleft}


\end{document}